\documentclass[aps,10pt,prd,notitlepage,showpacs,nofootinbib,superscriptaddress, compressed]{revtex4-1}

\pdfoutput=1 
\usepackage[utf8]{inputenc}
\usepackage[english]{babel}
\usepackage{amsmath}
\usepackage{graphicx}
\usepackage{dcolumn}
\usepackage{multirow}
\usepackage{pbox}
\usepackage{amssymb}
\usepackage{epsfig}
\usepackage[dvipsnames]{xcolor}
\usepackage{slashed}
\usepackage{amssymb}
\usepackage{ mathrsfs }
\usepackage{color}
\usepackage[font=small]{caption}
\usepackage[font=small]{subcaption}
\usepackage{url}

\usepackage{booktabs}
\graphicspath{{pic/}}

\definecolor{DarkBlue}{rgb}{0.7, 0.4, 1} 
\definecolor{Blue}{rgb}{0, 0.8, 0} 
\definecolor{MyLightBlue}{rgb}{0.5,0.7,1.9}
\definecolor{MyGreen}{rgb}{0.0,0.2, 0.0}
\definecolor{MyBrickRed}{rgb}{0, 0.5, 0.2}
\RequirePackage{hyperref}
\hypersetup{colorlinks, citecolor=Blue,linkcolor=DarkBlue, urlcolor=Green}

\newcommand {\ignore}[1]{}

\newcommand{\bea}{\begin{eqnarray}}
\newcommand{\eea}{\end{eqnarray}}

\makeatletter

\makeatletter
\renewcommand\@makecaption[2]{%
  \par
  \vskip\abovecaptionskip
  \begingroup
  
   \small\rmfamily
    \begingroup
     \samepage
     \flushing
     \let\footnote\@footnotemark@gobble
     \@make@capt@title{#1}{#2}\par
    \endgroup
  \endgroup
  \vskip\belowcaptionskip
}
\makeatother

\DeclareUnicodeCharacter{2212}{-}
\begin{document}
\title{Testing tree level TeV scale seesaw scenarios in $\mu$TRISTAN }
\author{Arindam Das}
\email{adas@particle.sci.hokudai.ac.jp}
\affiliation{Institute for the Advancement of Higher Education, Hokkaido University, Sapporo 060-0817, Japan}
\affiliation{Department of Physics, Hokkaido University, Sapporo 060-0810, Japan}
\author{Jinmian Li}
\email{jmli@scu.edu.cn}
\affiliation{College of Physics, Sichuan University, Chengdu 610065, China}
\author{Sanjoy Mandal}
\email{smandal@kias.re.kr}
\affiliation{Korea Institute for Advanced Study, Seoul 02455, Korea} 
\author{Takaaki Nomura}
\email{nomura@scu.edu.cn}
\affiliation{College of Physics, Sichuan University, Chengdu 610065, China}
\author{Rao Zhang}
\email{zhangrao@stu.scu.edu.cn}
\affiliation{College of Physics, Sichuan University, Chengdu 610065, China}
\begin{abstract}
We investigate TeV scale seesaw scenarios at $\mu^+ e^-$ and $\mu^+ \mu^+$ colliders in the $\mu$TRISTAN experiment. In minimal type-I seesaw scenario we consider two generations of Standard Model (SM) singlet heavy Majorana type Right Handed Neutrinos (RHNs) which couples with SM gauge bosons through light-heavy neutrino mixing. We discuss the prospects of probing heavy neutrinos via the processes such as $\mu^+e^-\to \nu N_i\to e^+ j j\nu$ or $\mu^- j j\nu$ for $\sqrt{s}=346$~GeV and $1\text{ ab}^{-1}$ luminosity. Studying these process, we estimate limits on the light-heavy neutrino mixing angles as a function of heavy neutrino mass, which could be two orders of magnitude stronger than electroweak precision data. Further, we study the effect of doubly charged scalar boson $(H^{++})$ from the type-II seesaw scenario in $\mu^+ \mu^+$ collision at $\sqrt{s}=2$ TeV. In this case we consider $\mu^+ \mu^+ \to \ell_i^+ \ell_j^+$ and $\mu^+ \mu^+ \to H^{++} Z/ \gamma$ processes followed by the same sign dilepton decay of $H^{++}$. We find that events involving $e^+ e^+$ among these final states can probe the neutrino mass ordering in $\mu$TRISTAN experiment at 5$\sigma$ significance. In addition to that we study the production of positively charged triplet fermion in $\mu$TRISTAN following $\mu^+ \mu^+ \to \mu^+ \Sigma^+$ process where $\Sigma^+$ decays into $\mu^+ jj$ mode through $Z$ boson exchange. Considering a triplet at 1 TeV and studying SM backgrounds we estimate the discovery potential of $\mu^+ \mu^+ jj$ signal at $\mu$TRISTAN with respect to projected luminosity.
\noindent 
\end{abstract}

\maketitle

\section{Introduction}
The origin of tiny neutrino mass and flavor mixing \cite{ParticleDataGroup:2020ssz} is an unknown, long standing puzzle as Standard Model (SM) fails to explain it. It leads us to extend SM in a variety of ways to explain such interesting observation. Simply a dimension five operator \cite{Weinberg:1979sa} within SM opened the door to incorporate beyond the Standard Model (BSM) physics through particle extension of the SM. 
A fascinating approach to study such anomalies is to introduce SM-singlet heavy Majorana type Right Handed Neutrinos (RHNs) to generate tiny neutrino mass through suppression 
of a lepton number violating mass scale \cite{Schechter:1980gr,Minkowski:1977sc,Gell-Mann:1979vob,Yanagida:1979as,Sawada:1979dis,Mohapatra:1979ia}- commonly recognized as type-I seesaw scenario. SM-singlet RHNs mix with the SM light neutrinos to interact with the SM gauge bosons and Higgs. However, RHNs are hitherto unidentified and considered that it could have mass from a light scale up to a very heavy scale depending on its role in a variety of theoretical aspects \cite{Das:2012ze,Deppisch:2015qwa,Das:2018hph,Das:2017nvm,Bolton:2019pcu, Coloma:2020lgy,Denton:2021czb,Dasgupta:2021ies,Ballett:2019bgd,Carbajal:2022zlp,Tastet:2020tzh,Abada:2016plb,Chun:2019nwi}. Recently LHC is also looking for prompt RHNs from same sign dilepton plus dijet and trilepton plus missing energy modes respectively to provide limits on the RHN mass and its mixing with light neutrinos \cite{CMS:2018iaf,CMS:2018jxx,CMS:2022hvh,CMS:2022fut}. 

Apart from the singlet fermion extension of the SM, there is another simple but interesting aspect where SM is extended by an $SU(2)$ triplet scalar with hypercharge $Y=+2$ arising as an UV-complete scenario of the dimension five operator. This scenario is called the type-II seesaw scenario \cite{Schechter:1981cv,Magg:1980ut,Cheng:1980qt,Lazarides:1980nt,Mohapatra:1980yp, 
Das:2024yvt,Mandal:2022zmy,Mandal:2022ysp}. Through the inclusion of the triplet scalar with small vacuum expectation value (VEV) $v_\Delta$, Majorana type tiny neutrino mass in $\mathcal{O}$(eV) scale can be generated at the tree level followed by flavor mixing. Such a scenario could evolve large neutrino Yukawa coupling of $\mathcal{O}$(1). Due to the gauge structure, the triplet (complex) scalar interacts with the SM gauge bosons, SM lepton and Higgs doublet developing a variety of decay modes of the triplet scalar~\cite{deSalas:2020pgw,10.5281/zenodo.4726908}. As a result such triplet scalars can be tested at high energy colliders following different decay modes of its doubly, singly charged and neutral scalar multiplets \cite{DELPHI:2002bkf,ATLAS:2017xqs,CMS:2017pet,ATLAS:2018ceg,CMS:2014mra,Antusch:2018svb,DeBlas:2019qco,Ashanujjaman:2022tdn,Ashanujjaman:2022ofg,Ashanujjaman:2021txz}. We point out that for $v_\Delta \leq 10^{-4}$, doubly charged Higgs dominantly decays into same sign dilepton mode whereas for $v_\Delta > 10^{-4}$, it decays into same sign $W$ bosons. In the first case, doubly charged scalar mass below 1080 GeV \cite{ATLAS:2022pbd}
was not observed whereas in the second case the upper limit is around 350 GeV \cite{ATLAS:2021jol}. The singly charged and neutral multiplets from the type-II seesaw scenario have been tested at the LHC from gluon-fusion channels \cite{CMS:2015lsf,ATLAS:2018gfm,ATLAS:2018ntn,ATLAS:2017eiz,CMS:2018rmh,ATLAS:2018sbw}. These processes are suppressed by an additional factor of $\mathcal{O}(v_\Delta^2)$ making them irrelevant for type-II seesaw scenario. 

Type-III seesaw can be considered as another interesting framework which is realized extending the SM by an $SU(2)_L$ triplet fermion with zero hypercharge~\cite{Foot:1988aq}. The triplet fermion consists of both singly charged and charge-neutral multiplets. The mass of the light neutrino is generated by the VEV of the charge-neutral multiplet following the seesaw mechanism. These charge-neutral multiplets can mix with the SM neutrinos, enabling their interaction with the SM bosons. The Type-III seesaw scenario has been explored at high-energy colliders to investigate various phenomenological aspects, including both prompt and displaced signatures from multi-lepton and multi-jet channels~\cite{ATLAS:2022yhd,CMS:2022nty,Das:2020uer,Das:2020gnt,Ashanujjaman:2021jhi,Ashanujjaman:2021zrh}, setting a limit on their masses around 1 TeV.

 In this paper we study the tree level minimal seesaw models at  muti-TeV same sign positively charged muon collider or positively charged muon-electron collider~\cite{Heusch:1995yw,Belanger:1995nh,Gluza:1995ky,Raidal:1997tb,Shiltsev:2010qg,Rodejohann:2010jh,Arbuzov:2021oxs,Bondarenko:2021eni,Han:2021nod,Hamada:2022uyn,Hamada:2022mua,Fridell:2023gjx,Lichtenstein:2023iut,Mekala:2023kzo,Mekala:2023diu,Dev:2023nha,Calibbi:2024rcm,Bai:2024skk} at $\mu$TRISTAN experiment using ultra-cold anti-muon technology from J-PARC~\cite{Abe:2019thb}. In this experiment $\mu^+ e^-$ collision could occur at $\sqrt{s}=346$ GeV where $\mu^+$ energy is $E_\mu=1$ TeV and $e^-$ energy is $E_e=30$ GeV $(\sqrt{s}=2\sqrt{E_\mu E_e})$ at 1 ab$^{-1}$ luminosity. This could be upgraded to $\mu^+ \mu^+$ collider at $\sqrt{s}=2$ TeV or higher where muon energy is considered to be $1$ TeV with $100$ fb$^{-1}$ luminosity or higher. Such a machine can potentially serve as a Higgs factory and participate in various phenomenological aspects \cite{Bossi:2020yne,Lu:2020dkx,Cheung:2021iev,Yang:2020rjt,Liu:2021jyc,Yang:2023ojm,
Das:2022mmh} involving physics beyond the SM.

The unique initial state of the $\mu$TRISTAN experiment allows us to study the tree level seesaw scenarios from interesting perspectives. In case of $\mu^+ e^-$ collision we consider the heavy neutrino production from the type-I seesaw where only a $t-$channel $W$ mediated scenario will appear provided the RHN mass $(M_N)$ resides in the electroweak or TeV scale. Due to the light-heavy mixing, this process will be suppressed by modulus square of the respective mixing. We consider that there is at least one  RHN which reside at the electroweak scale and can be produced at the colliders. 
In this collider we can explore this RHN and its corresponding mixing with electron, muon and tau. The RHN produced in this scenario can dominantly decay into an electron,  muon or tau and $W$ boson followed by the hadronic decay of its daughter $W$ boson. Finally from the lepton and jets it could be possible to reconstruct the RHN. Hence we can estimate the limits on mass and mixings of RHN studying signal and corresponding SM backgrounds.  For $\mu^+ e^-$ collision we consider $45$ GeV $\leq M_{N} \leq 340$ GeV.  In case of $\mu^+ \mu^+$ collider we can study a unique behavior of the RHNs where a $t-$channel RHN mediated process could produce a pair of same sign $W$ boson reflecting the Majorana nature of the heavy neutrino. However, this process is suppressed by the fourth power of the modulus of the mixing. 

In case of type-II seesaw scenario, $\mu^+\mu^+$ collision could play a crucial role to probe the effect of doubly charged scalar multiplet. In this case we consider two scenarios. In the first case we consider the doubly charged scalar mediated same-sign dilepton production scenario in the $s-$channel process accompanied by the $t-$channel same sign dilpeton production mode mediated by SM neutral bosons. Here we prefer to consider non-muonic final states to reduce SM backgrounds. Followed by this scenario, we study the production of doubly charged scalar in association with $Z$ boson or photon. If $Z$ boson is produced, then it is considered to decay into dominant mode of jets resulting a final state of same sign dilepton and two jets where same sign dileptons are produced from the doubly charged scalar. If the doubly charged scalar is produced with a photon, the final state consists of a same sign dilepton from doubly charged scalar plus a photon. It's important to note that the dilepton decay of a doubly charged Higgs comes from the Yukawa coupling between the lepton doublet and scalar triplet, which is determined by the oscillation parameters, particularly the neutrino mass ordering. Therefore, studying the above mentioned processes could provide insights into the ordering of light neutrino masses, for example, normal ordering (NO) and inverted ordering (IO). Finally, we compare the left-right asymmetry $(\mathcal{A}_{\rm{LR}})$ for the SM and doubly charged scalar induced dilepton production process. 

In type-III seesaw, one can have production channels such as $\mu^+e^-\to\Sigma^+\Sigma^-$ and $\mu^+\mu^+\to \Sigma^+\Sigma^+$ at $\mu$TRISTAN.  But due to the current bound on the triplet fermion mass, type-III seesaw is not suitable to study at $\mu$TRISTAN. In addition to that, these processes will be suppressed by fourth power of the light-heavy mixing making the scenario less interesting. However, there is an interesting scenario which could be studied in the context of $\mu^+ \mu^+$ collider. In this case singly charged multiplet of the triplet fermion could be produced in association with a positively charged muon following $\mu^+ \mu^+ \to \Sigma^+ \mu^+$ and $\Sigma^+$ will decay into $\mu^+ Z$ followed by hadronic decay of $Z$ boson. Hence a same sign two muon final state in association with two jets $\mu^+ \mu^+ jj$ could be studied at $\sqrt{s}=2$ TeV while $M_{\Sigma}=1$ TeV.

The paper is arranged in the following way. In Sec.~\ref{model} we briefly discuss the type-I and type-II seesaw models at the tree level which could be tested at $\mu$TRISTAN experiment. We give detailed analysis on the BSM searches in Sec.~\ref{res} followed by discussions on limits obtained on the light-heavy neutrino mixings. Finally we conclude our article in Sec.~\ref{conc}.
\section{Testable seesaw models at the tree level}
\label{model}
Among many proposals of neutrino mass generation mechanism, we discuss neutrino mass generation mechanism at the tree level involving type-I, type-II and type-III seesaw models which could reproduce the neutrino oscillation data and flavor mixing.
\subsection{Type-I seesaw}
In type-I seesaw, SM-singlet RHNs $(N_R^{\beta})$ are introduced to extend the SM which couples with SM lepton doublet $(\ell^{\alpha})$ and Higgs doublet $(\Phi)$. We can then write down the relevant part of the Lagrangian as 
\bea
\mathcal{L}_{\rm int} \supset -y_D^{\alpha\beta} \overline{\ell_L^{\alpha}}\Phi N_R^{\beta} 
                   -\frac{1}{2} M_N^{\alpha \beta} \overline{N_R^{\alpha C}} N_R^{\beta}  + H. c. .
\label{typeI}
\eea
where $\alpha$ and $\beta$ stand for the generation indices. The second term in Eq.~\ref{typeI} is the lepton number violating Majorana mass term for the RHNs. After the electroweak symmetry is broken by the generation of VEV of the SM Higgs $v_\Phi$, one obtain the Dirac mass term of the neutrinos as $M_{D}= \frac{y_D v_\Phi}{\sqrt{2}}$. Hence we can write neutrino mass matrix incorporating the Dirac and Majorana masses in the following as 
\bea
M_{\nu}=\begin{pmatrix}
0&&M_{D}\\
M_{D}^{T}&&M_N
\end{pmatrix}.
\label{typeInu}
\eea
Now diagonalizing Eq.~\ref{typeInu}, the seesaw formula for the light Majorana neutrinos mass can be obtained as 
\bea
m_{\nu} \simeq - M_{D} M_N^{-1} M_{D}^{T}.
\label{seesawI}
\eea
To obtain a light neutrino mass of $\mathcal{O}(0.1)$ eV from a heavy RHN of mass 
$M_N$ around $\mathcal{O}(100)$ GeV, the Dirac Yukawa coupling $y_D$ could be $\mathcal{O}(10^{-6})$, however, the $Y_D$ could reach at $\mathcal{O}(1)$ if general parametrization for the seesaw scenario is considered \cite{Casas:2001sr}. This scenario is considered in our article. We assume, $M_D M_N^{-1} \ll 1$ so that the light neutrino flavor eigenstates ($\nu$) can be written in terms of  the mass eigenstates of light neutrino ($\nu_m$) and heavy neutrino ($N_m$) as $\nu \simeq {\cal N} \nu_m + V N_m$ where $V = M_D M_N^{-1}$ and ${\cal N} =  \left(1 - \frac{1}{2} \epsilon \right) U_{\rm PMNS}$ with $\epsilon = V^* V^T$ and $U_{\rm PMNS}$ is the usual neutrino mixing matrix applied to diagonalize $m_\nu$ in the following way   
\bea
   U_{\rm PMNS}^T m_\nu U_{\rm PMNS} = {\rm diag}(m_1, m_2, m_3). 
   \label{eigen}
\eea
In presence of the parameter $\epsilon$, mixing matrix ${\cal N}$ is non-unitary \cite{Antusch:2006vwa,Abada:2007ux,Antusch:2014woa}. Now we replace neutrino flavor eigenstates in charged current (CC) and neutral current (NC) interactions of the SM and obtain 
\bea 
\mathcal{L}_{\rm CC}= 
 -\frac{g}{\sqrt{2}} W_{\mu}
  \overline{e} \gamma^{\mu} P_L 
   \left( {\cal N} \nu_m+  V N_m \right) + \text{h.c.}, 
\label{CC}
\eea
where $e$ denotes the three generations of the charged leptons in the vector form and 
\bea 
\mathcal{L}_{\rm NC}= 
 -\frac{g}{2 c_w}  Z_{\mu} 
\left[ 
  \overline{\nu_m} \gamma^{\mu} P_L ({\cal N}^\dagger {\cal N}) \nu_m 
 +  \overline{N_m} \gamma^{\mu} P_L ( V^\dagger  V) N_m 
+ \left\{ 
  \overline{\nu_m} \gamma^{\mu} P_L ({\cal N}^\dagger   V) N_m 
  + h.c. \right\} 
\right] , 
\label{NC}
\eea
where $c_w=\cos \theta_w$ is the weak mixing angle $P_{\rm L(R)} =\frac{1}{2} (1\mp \gamma_5)$. 
\ignore{Because of non-unitarity ${\cal N}$, the flavor-changing neutral current could occur through the first term of Eq.~(\ref{NC}) involving ${\cal N}^\dagger {\cal N}$ which is not equal to unity.} 
In our scenario RHN mass is a free parameter. If  RHN mass lies between $10$ GeV $\leq M_N < M_W$, RHN undergoes three-body decay via offshell $W,Z$ and Higgs boson. 
If the RHNs are heavier than the SM gauge and scalar bosons then dominant two-body decay modes are $N \to \ell W$, $\nu_{\ell} Z$, $\nu_{\ell} h$, respectively.
\ignore{and the corresponding partial decay widths can be given by
\bea
\Gamma(N_{i} \rightarrow \ell_\alpha W) 
 &=& \frac{g^2 |V_{i \alpha}|^{2}}{64 \pi} 
 \frac{ (M_N^2 - M_W^2)^2 (M_N^2+2 M_W^2)}{M_N^3 M_W^2} ,
\nonumber \\
\Gamma(N_{i} \rightarrow \nu_\alpha Z) 
 &=& \frac{g^2 |V_{i \alpha}|^{2}}{128 \pi c_w^2} 
 \frac{ (M_N^2 - M_Z^2)^2 (M_N^2+2 M_Z^2)}{M_N^3 M_Z^2} ,
\nonumber \\
\Gamma(N_{i} \rightarrow \nu_\alpha h) 
 &=& \frac{ |V_{i \alpha}|^2 (M_N^2-M_h^2)^2}{32 \pi M_N} 
 \left( \frac{1}{v }\right)^2 ,
\label{widths}
\eea 
respectively. The partial decay width of the heavy RHN into $W^\pm$ being twice as large as into a neutral one due to the two degrees of freedom of $W^{\pm}$.}
\begin{figure}[h]
\centering
\includegraphics[width=0.5\textwidth]{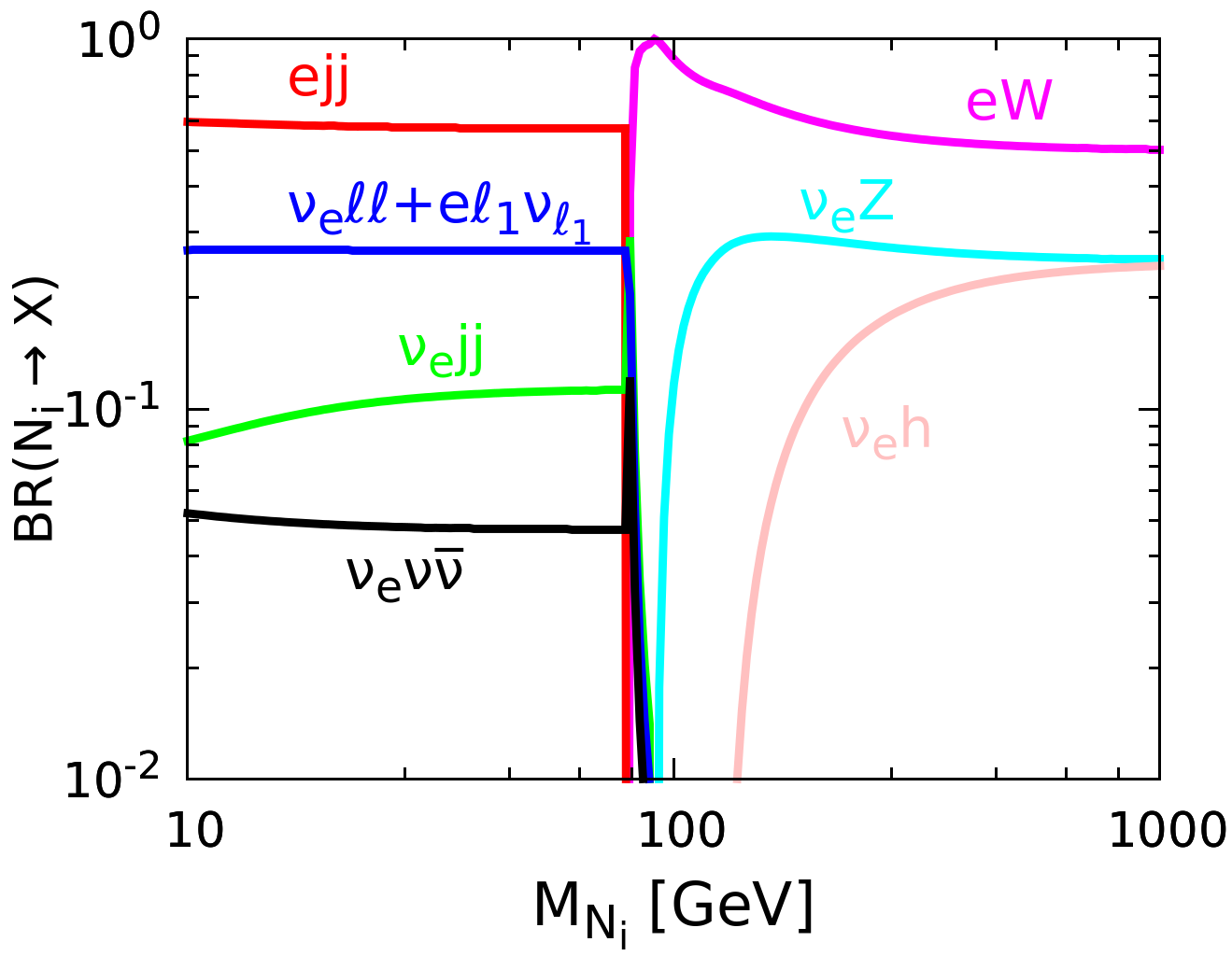}
\caption{Branching ratios of $N_i$ to different final states under the assumption $V_{eN_i}\neq 0, V_{\mu N_i}=0$ and $V_{\tau N_i}=0$. We show the branching ratios to three body leptonic channels $e\ell_1\nu_{\ell_1}+\nu_e\ell\ell$, $\nu_e\nu\bar{\nu}$ and semi-leptonic channels such as $ejj$ and $\nu_e jj$, where $\ell=e,\mu,\tau$, $\ell_1=\mu,\tau$. For relatively large $M_{N_i}$ the two body decay channels such as $eW$, $\nu_e Z$ and $\nu_e h$ dominates.}
\label{fig1}
\end{figure}
In Fig.~\ref{fig1}, we show the branching ratios of heavy neutrino $N$ to various final states for $10$ GeV $\leq M_N \leq 1000$ GeV assuming that $V_{eN}\neq 0, V_{\mu N}=0$ and $V_{\tau N}=0$. 
\ignore{We point out that similar behavior will be observed with the choice $V_{eN_i}=0, V_{\mu N_i}=1$ and $V_{\tau N_i}=0$. This is minimal scenario in case of type-I seesaw.}
For RHN lighter than $W$ boson $(M_{N}< M_W)$ the dominant modes involve three body leptonic or semi-leptonic final states, whereas for RHN heavier than $W$ boson $(M_{N}>M_W)$, two body decay modes dominate. Note that for very heavy RHN where $M_N \geq \mathcal{O}(1)$ TeV with non-zero light-heavy mixing, the branching ratios can be obtained as
\begin{align}
\text{BR}(N\to\ell W):\text{BR}(N\to\nu_{\ell}Z):\text{BR}(N\to\nu_{\ell}h)=2:1:1.
\label{eq:gg}
\end{align} 
The above asymptotic behavior of heavy neutrino decay branching ratios is due to the Goldstone equivalence theorem. Note that for sufficiently small mixing and relatively small $M_{N}$, heavy RHNs could be long-lived candidates which that can travel a long distances before their decay into different daughter particles. This scenario gives rise to displaced vertex scenarios. We consider $M_N$ and $|V_{\ell N}|^2$ as free parameters and $V_{\ell N}$ is always large enough so that RHNs suffer prompt decay.
\subsection{Type-II seesaw scenario}
In case of type-II seesaw scenario, SM is extended by an $SU(2)_L$ triplet scalar $\Delta=(\Delta^{++},\Delta^{+},\Delta^{0})^{T}$. The triplet scalar is described in the matrix form with the help of Pauli spin matrices as 
\begin{equation}
\Delta=\left(\begin{array}{cc}
\Delta^{+}/\sqrt{2} & \Delta^{++}\\
\Delta^{0} & -\Delta^{+}/\sqrt{2}
\end{array}\right),
\end{equation}
and due to its $SU(2)_L$ charge, it interacts with the SM gauge bosons. In addition to that it participates in a Yukawa coupling with SM lepton doublets. The corresponding Lagrangian with the scalar kinetic term can be written as
\begin{equation}
{\cal L}_{{\rm type\ II}}= \left[iY_{\Delta\alpha\beta}L_{\alpha}^{T}C^{-1}
    \tau_2\Delta L_{\beta}+\text{h.c.}\right]+\left(D_{\mu}\Phi\right)^{\dagger}\left(D^{\mu}\Phi\right) +\left(D_{\mu}\Delta\right)^{\dagger}\left(D^{\mu}\Delta\right)-V(\Phi,\Delta),
    \label{type2}
\end{equation}
where $Y_{\Delta}^{\alpha\beta}$ is the Yukawa coupling and it is a symmetric complex matrix. Here $L_\alpha$ represents SM lepton doublets, $C$ stands for the charge conjugation operator, $D_\mu$ shows covariant derivative of the related scalar field and 
$V(\Phi,\Delta)$ is the scalar potential which could be expressed as
\begin{equation}
\begin{aligned}
 V(\Phi,\Delta) = & -m_{\Phi}^{2}\Phi^{\dagger}\Phi + \frac{\lambda}{4}(\Phi^{\dagger}\Phi)^{2} +  \tilde{M}_{\Delta}^{2}{\rm Tr}\left[\Delta^{\dagger}\Delta\right]+\lambda_{2}\left[{\rm Tr}\Delta^{\dagger}\Delta\right]^{2}+\lambda_{3}{\rm Tr}\left[\Delta^{\dagger}\Delta\right]^{2}\\
  &+ \left[\mu \Phi^{T}i\sigma_{2}\Delta^{\dagger}\Phi+\text{h.c.}\right]+\lambda_{1}(\Phi^{\dagger}\Phi){\rm Tr}\left[\Delta^{\dagger}\Delta\right]+\lambda_{4}\Phi^{\dagger}\Delta\Delta^{\dagger}\Phi.\label{eq:potential}
\end{aligned}
\end{equation}
In this case Higgs triplet UV-completion of the Weinberg operator could be characterized by the induction of a tiny VEV of $\Delta$. The lepton number violation $\mu$-term in Eq.~\ref{eq:potential} induces a small VEV of triplet scalar as
\begin{equation}
	v_\Delta\approx\frac{\mu v_\Phi^2}{\sqrt{2}\tilde{M}^2_{\Delta}},
	\label{Eq:induced_vev}
\end{equation}
where we see that the smallness of $v_\Delta$ can be realised by either assuming a small $\mu$ or, a large value for $\tilde{M}_{\Delta}$ characterizing the triplet scalar mass~\cite{Mandal:2022zmy,Mandal:2022ysp}. The triplet VEV $v_\Delta$ could be constrained from the $\rho$ parameter leading to upper bound of of the triplet VEV as $\mathcal{O}(1\text{ GeV})$~\cite{ParticleDataGroup:2020ssz}.
After the electroweak symmetry breaking seven out of them have definite mass being evolved as physical fields. These scalar fields can be written as  $H^{\pm\pm}$, $H^{\pm}$ for the charged multiplets; $h$, $H^0$ and $A^0$ as neutral scalar fields respectively. Here $H^{\pm\pm}$ is simply $\Delta^{\pm\pm}$ as presented in the triplet scalar field $\Delta$. Using the fact $v_\Delta\ll v_\Phi$, the masses of physical Higgs bosons can be approximated as
\begin{align}
m_{H^{\pm\pm}}^2\simeq M_\Delta^2-\frac{\lambda_4}{2}v_\Phi^2,\,\,\,\, m_{H^\pm}^2\simeq M_\Delta^2-\frac{\lambda_4}{4}v_\Phi^2,\,\,\,\, m_h^2\simeq 2\lambda v_\Phi^2\,\,\,\,\text{and}\,\,\,\,m_{H^0}^2\approx m_{A^0}^2\simeq M_{\Delta}^2 ,
\end{align}
so their mass differences can be written as
\begin{align}
\Delta m\equiv m_{H^0/A^0}-m_{H^\pm}\equiv m_{H^\pm}-m_{H^{\pm\pm}}\approx \frac{\lambda_4}{8}\frac{v_\Phi^2}{M_\Delta}.
\end{align}
Hence we could express all the physical Higgs masses in terms of $m_{H^{\pm\pm}}$ and $\Delta m$
only. The mass splitting between singly and doubly charged multiplets of the triplet scalar affects $S, T$ and $U$ parameters putting a stringent constraint as $|\Delta m|\lesssim 40$ GeV~\cite{Aoki:2012jj}. Depending on the value and sign of $\lambda_4$ we obtain three different spectra: (i) $\lambda_4=0:\,\,\Delta m\approx 0\,( m_{H^{\pm\pm}}\simeq m_{H^\pm}\simeq m_{H^0/A^0})$, (ii) $\lambda_4<0:\,\,\Delta m<0\,( m_{H^{\pm\pm}} > m_{H^\pm} > m_{H^0/A^0})$ and (iii) $\lambda_4>0:\,\,\Delta m>0\,( m_{H^{\pm\pm}} < m_{H^\pm} < m_{H^0/A^0})$, respectively. The triplet scalar interacts with the SM gauge bosons and fermions.
\par Through the induced triplet VEV $v_\Delta$, the Yukawa term in Eq.~\ref{type2} generate the following Majorana mass term for light neutrinos
\begin{align}
\mathcal{L}_{\rm Majorana}=\frac{1}{2}\overline{\nu_{\alpha L}^c} m_{\alpha\beta}^{\nu}\nu_{\beta L} + \text{H.c.} \,\,\, \text{with} \,\,\,  m^\nu_{\alpha \beta} = \sqrt{2} (Y_\Delta)_{\alpha \beta} v_\Delta .
\label{eq:numass}
\end{align} 
From Eq.~(\ref{eq:numass}) we find that neutrino mass is directly proportional to the triplet VEV. Therefore tiny neutrino mass can be ensured by small triplet VEV which could be satisfied either by a small trilinear term $\mu$ or being suppressed by a large $\tilde{M}_{\Delta}$. 
Using neutrino oscillation data and $U_{\rm PMNS}^T m_\nu U_{\rm PMNS} =  \text{diag}(m_1, m_2, m_3)$ to diagonalize the neutrino mass matrix we express 
\bea
Y_\Delta = \frac{1}{\sqrt{2 v_\Delta}}\, U_{\rm{PMNS}} \,m_\nu^{\rm{diag}} \, U_{\rm{PMNS}}^T,
\label{eq:Yukawa}
\eea
where $m_{\nu}^{\rm{diag}}= \text{diag}(m_1, m_2, m_3)$ are the neutrino mass eigenvalues which carry the information of neutrino oscillation for NO and IO along with $U_{\rm{PMNS}}$ which is the PMNS matrix. 

\begin{figure}[]
\centering
\includegraphics[width=0.49\textwidth]{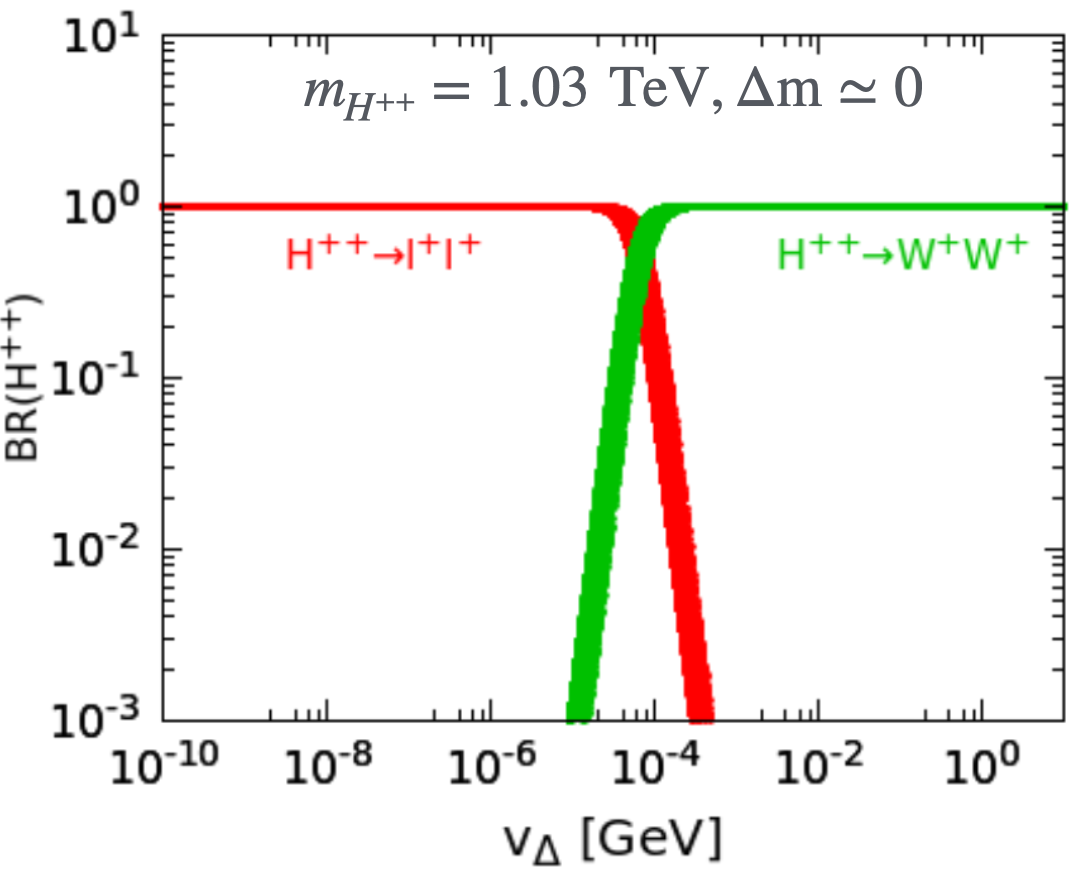}
\includegraphics[width=0.49\textwidth]{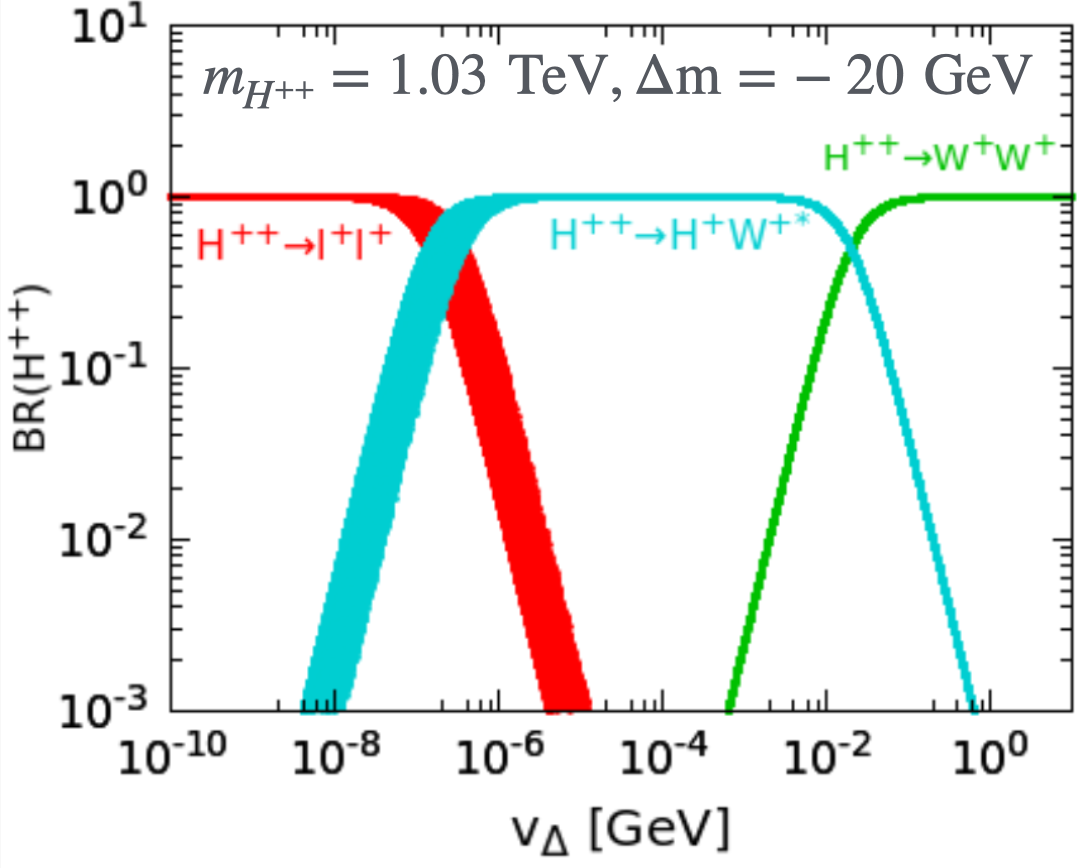}
\caption{Branching ratios of $H^{\pm\pm}$ for $m_{H^{\pm\pm}}=1.03$~TeV with $\Delta m\approx 0$~(left panel) and $\Delta m=-20$~GeV~(right panel), respectively. We consider three decay mode of $H^{\pm\pm}$ into $\ell^{\pm}\ell^{\pm}$ (red), $W^{\pm} W^{\pm}$ (green) and $ H^{\pm}W^{\pm *}$ (cyan), respectively. The decay mode $H^{\pm\pm}\to H^{\pm}W^{\pm *}$ is only open for $\Delta m < 0$ and further dominates in the region of intermediate $v_\Delta$ when $\Delta m$ is relatively large. }
\label{fig2}
\end{figure}

In Fig.~\ref{fig2}, we show the branching ratios of $H^{\pm\pm}$ for $m_{H^{\pm\pm}}=1.03$~TeV with $\Delta m\approx 0$~(left panel) and $\Delta m=-20$~GeV~(right panel), respectively. For small $\Delta m$, depending on $v_\Delta$, $H^{\pm\pm}$ dominantly decays either into same-sign dileptons for $v_\Delta\leq 10^{-4}$~GeV or gauge bosons for $v_\Delta > 10^{-4}$~GeV. On the other hand from the right panel of Fig.~\ref{fig2} with $\Delta m=-20$ GeV, the cascade mode $H^{\pm\pm}\to H^{\pm} W^{\pm *}$ dominate over the leptonic and diboson decay modes for an intermediate $v_\Delta$. The doubly charged multiplet of the triplet scalar provide tightest constraints from the searches at the LHC. The collider searches strictly depend on $v_\Delta$ and $\Delta m$ which governs the decay modes of the doubly charged multiplets, see Fig.~\ref{fig2}. For our study we consider a degenerate mass spectrum $m_{H^{\pm\pm}} = m_{H^\pm} = m_{H^0} = m_{A^0}$ and the triplet VEV in the region $v_\Delta < 10^{-4}$~GeV, such that $H^{\pm\pm}\to\ell_i^\pm\ell_j^\pm$ is the dominant decay mode. In this case, LHC has ruled out doubly charged scalar mass below 1080 GeV \cite{ATLAS:2022pbd}. 
\subsection{Type-III seesaw scenario}
\begin{figure}[]
\centering
\includegraphics[width=0.46\textwidth]{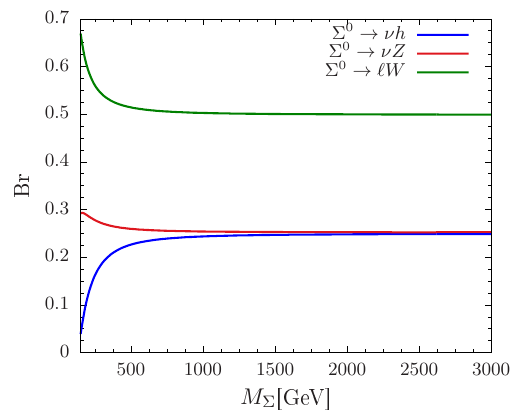}
\includegraphics[width=0.46\textwidth]{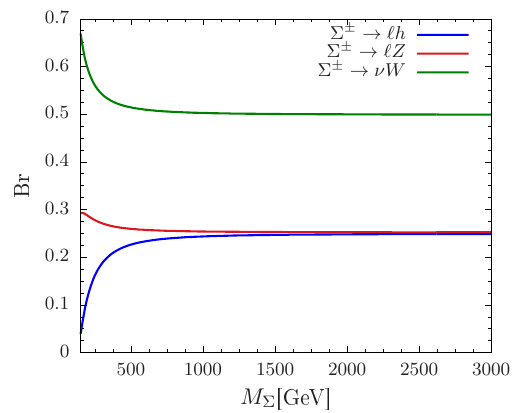}\\
\caption{Branching ratio (Br) of $\Sigma^0$~(left) and $\Sigma^{\pm}$~(right) into the SM particles as a function of $M_\Sigma$ for $|V_{\mu \Sigma}|^2=2.89 \times 10^{-4}$ considering $V_{e\Sigma}= V_{\tau \Sigma}=0$ from \cite{delAguila:2008pw,delAguila:2008cj}.}
\label{branching ratio1}
\end{figure}

n the type-III seesaw model SM is extended by three generations of an $SU(2)_L$ triplet fermion $(\Psi)$ with zero hypercharge~\cite{Foot:1988aq}. Inclusion of such triplets helps the generation of nonzero but tiny neutrino mass through the seesaw mechanism. The Lagrangian can be written as
\begin{align}
\mathcal{L}=\mathcal{L}_{\text{SM}}+ \text{Tr}(\overline{\Psi}i \gamma^\mu D_\mu \Psi)-\frac{1}{2}M \text{Tr}(\overline{\Psi}\Psi^c+\overline{\Psi^c}\Psi)-\sqrt{2}(\overline{\ell_L}Y_D^\dagger \Psi H + H^\dagger \overline{\Psi} Y_D \ell_L)
\label{L}
\end{align}
where $D_\mu$ represents the covariant derivative, $M$ is the Majorana mass term. We consider three degenerate generation of the triplets. $Y_D$ is the Dirac Yukawa coupling between the SM lepton doublet $(\ell_L)$, SM Higgs doublet $(H)$ and the triplet fermion $(\Psi)$. For brevity, we have suppressed the generation indices. In this analysis we represent the relevant SM candidates, the triplet fermion and its charged conjugate $(\Psi^c = C\overline{\Psi}^T)$ as  in the following way 
\bea
\ell_L
 =
 \begin{pmatrix}
  \nu_{L}\\
  e_{L} \\
 \end{pmatrix},\,\,\,
H = 
 \begin{pmatrix}
 \phi^0\\
  \phi^-\\
 \end{pmatrix},\,\,\, 
\Psi=
 \begin{pmatrix}
  \Sigma^0/\sqrt{2}  &  \Sigma^+ \\
 \Sigma^-           &   -\Sigma^0/\sqrt{2}  \\
 \end{pmatrix}\,\,\text{and}\,\,
  \Psi^c=
 \begin{pmatrix}
  \Sigma^{0c}/\sqrt{2}  &  \Sigma^{-c} \\
 \Sigma^{+c}           &   -\Sigma^{0c}/\sqrt{2}  \\
 \end{pmatrix}.
 \label{L2}
\eea
To study the mixing between the SM charged leptons and $\Sigma^\pm$ we write the four degrees of freedom of each $\Sigma^\pm$ in terms of a Dirac spinor such as 
$\Sigma=\Sigma_R^-+\Sigma_R^{+c}$ where as $\Sigma^0$ are two component fermions with two degrees of freedom. The corresponding Lagrangian after the electroweak symmetry breaking can be written as 
\begin{align}
 -\mathcal{L}_{\text{mass}}=
  \begin{pmatrix}
  \overline{e}_L & \overline{\Sigma}_L \\
 \end{pmatrix}
 \begin{pmatrix}
  m_\ell & Y_D^\dagger v\\
  0 & M \\
 \end{pmatrix}
 \begin{pmatrix}
  e_R \\
  \Sigma_R \\
 \end{pmatrix} 
+
 \frac{1}{2}\begin{pmatrix}
  \overline{\nu_L^c} & \overline{\Sigma_R^0} \\
 \end{pmatrix}
 \begin{pmatrix}
  0 & Y_D^T \frac{v}{\sqrt{2}} \\
  Y_D\frac{v}{\sqrt{2}} & M \\
 \end{pmatrix}
 \begin{pmatrix}
  \nu_L \\
  \Sigma_R^{0c} \\
 \end{pmatrix}
 +h. c.
 \label{n1}
\end{align}
where $m_\ell$ is the Dirac type SM charged lepton mass. The $3\times3$ Dirac mass of the triplets can be written as $M_D=Y_D^T v/\sqrt{2}$. Diagonalizing the neutrino mass matrix in Eq.~(\ref{n1}) we can write the light neutrino mass eigenvalue as 
\bea
m_\nu \simeq -\frac{v^2}{2} Y_D^T M^{-1} Y_D = M_D M^{-1} M_D^{T}.
\label{n3}
\eea
The light neutrino flavor eigenstate can be expressed in terms of the light and heavy mass eigenstates in the following way
\bea
\nu \approx \nu_m + V \Sigma_m,
\label{n44}
\eea
where $\nu_m$ and $\Sigma_m$ represent the light and heavy mass eigenstates respectively where $V= M_D M^{-1}$. The relevant charged current, neutral current and higgs interaction involving SM leptons and triplet fermions can be found in Refs.~\cite{Abada:2007ux,Abada:2008ea,Biggio:2011ja,Das:2020uer}. 
Estimating partial decay widths  \cite{Das:2020uer} we show the branching ratios $\Sigma^{0,+}$ in Fig.~\ref{branching ratio1}. Note that when the triplet is very heavy, the branching ratios of $\Sigma^{0,+}$ to different mode follows the same relation as of Eq.~\ref{eq:gg} and this is again due to the Goldstone equivalence theorem.
In this analysis we consider the scenario where limit on the light-heavy mixing $|V_{\mu \Sigma}|^2 = 2.89\times 10^{-4}$ considering $V_{e\Sigma}= V_{\tau \Sigma}=0$ from EWPD following \cite{delAguila:2008pw,delAguila:2008cj}. We use this mixing as the benchmark limit in further analysis.
\section{Results and discussions}
\label{res}
In type-I seesaw scenarios, we investigate a single electron~(muon) final state in association with jets and neutrinos. In case of type-II seesaw scenario we study same-sign dilepton final state and doubly charged scalar multiplet production in association with $Z$ boson or photon. We also study left-right asymmetry form the same-sign dilepton mode. We study the production of positively charged multiplet of the triplet fermion at $\mu^+ \mu^+$ collider in association with $\mu^+$. Following the decay of the positively charged multiplet of the triplet we study same-sign dilepton and jets in final state from the type-III seesaw scenario. We describe complete search strategies and results in the following way: 
\subsection{Heavy neutrino search}
\begin{figure}[]
\centering
\includegraphics[width=1\textwidth,angle=0]{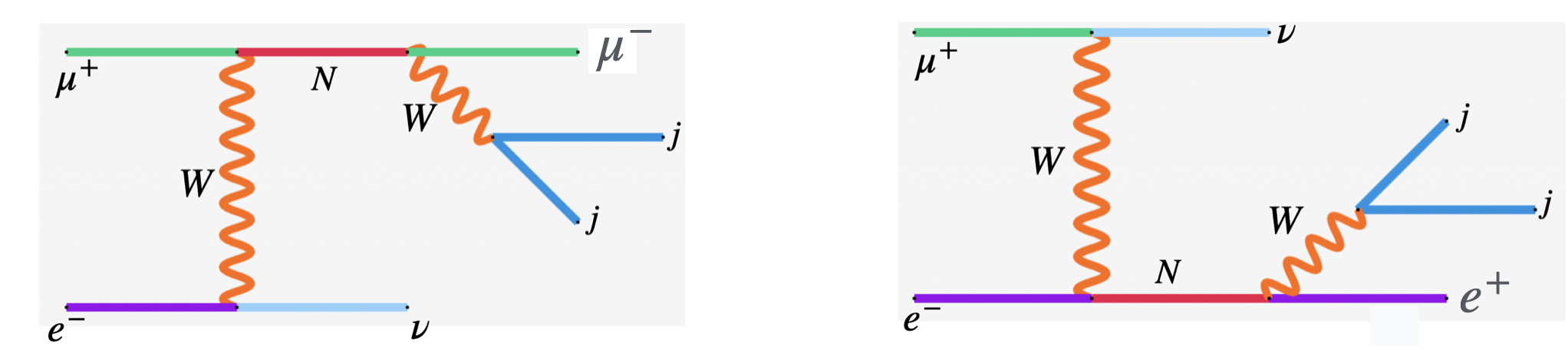}
\caption{Heavy neutrino production processes in $t-$channel and its dominant decay in $\mu$TRISTAN experiment. Following the heavy neutrino production we consider the heavy neutrino decays to either positron or muon in association with two jets.}
\label{fig3}
\end{figure}

\begin{figure}[]
\centering
\includegraphics[width=0.55\textwidth,angle=0]{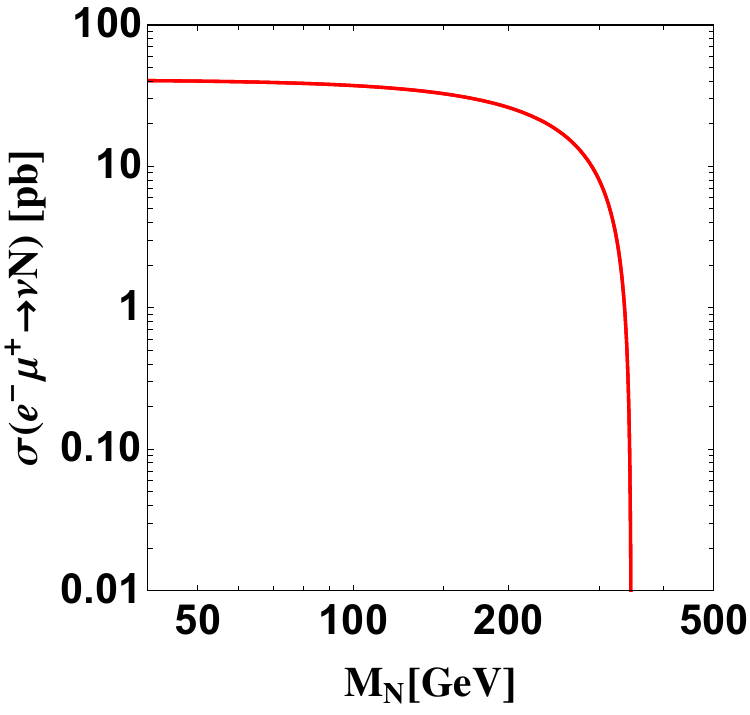}
\caption{Heavy neutrino production cross section as a function of heavy neutrino mass in $\mu$TRISTAN experiment at $\sqrt{s}=346$ GeV center of mass energy normalized by the light-heavy mixing squared.}
\label{fig4}
\end{figure}

At $\mu^+ e^-$ collider of the $\mu$TRISTAN experiment, the heavy neutrino can be produced as $\mu^+e^-\to\nu N$ through $W$ mediated $t-$channel process as shown in Fig.~\ref{fig3}.  The differential scattering cross section for this $t-$channel process can be given by
\bea
\frac{d\sigma(\mu^+ e^- \to N \nu)}{d\cos\theta}=(3.89\times 10^8~{\rm pb}) \times \frac{(s-M_N^2) }{32 \pi s^2} \times 
\Big[\frac{\frac{g^4}{4} \{(s- M_N^2)(1\pm\cos\theta)\times (\frac{1}{2} s-\frac{1}{4}(s -M_N^2)(1 \mp\cos\theta))\}}{(\frac{1}{2} (s- M_N^2)(1\mp\cos\theta)+M_W^2)^2+\Gamma_W^2 M_W^2}\Big],
\eea
where $g,\,M_W$ and $\Gamma_W$ are the $SU(2)$ gauge coupling, $W$ boson mass and the total decay width, respectively. To obtain the signal cross section, we integrated over the scattering angle $\theta$ in the range of $-1 \leq \cos \theta \leq 1$. Note that the production cross-section is normalized by the relevant light-heavy neutrino mixing square. The normalized production cross section in $\mu$TRISTAN experiment at $\sqrt{s}=346$ GeV center of mass energy are shown in Fig.~\ref{fig4}. We find that the production process $\mu^+ e^-\to N\nu$ followed by the decay chains such as $N\to\mu^-W^+/e^+ W^-$ and $W^\pm\to jj$ leads to collider signatures such as $\mu^-jj\nu$ or $e^+jj\nu$. 
For the signal, we only consider the final state with a hadronically decaying $W$ boson, because this channel suffers from lower background than the leptonic decaying $W$ channel at the lepton collider.
We find that final state $\mu^+ e^-\to N\nu\to \mu^- j j\nu$ can constrain either $|V_{\mu N}|^2$~(when $N$ is from $\mu^+$ beam) or $|V_{e N}^{*}V_{\mu N}|$~(when $N$ is from $e^-$ beam), whereas the final state $\mu^+ e^-\to N\nu\to e^+ j j\nu$ can constrain either $|V_{e N}|^2$~(when $N$ is from $e^-$ beam) or $|V_{e N}^{*}V_{\mu N}|$~(when $N$ is from $\mu^+$ beam). 
\par To study the signal and SM backgrounds, we simulate corresponding events using the event generation package \texttt{MadGraph5\_aMC@NLO}~\cite{Alwall:2014hca} using the UFO model files generated by \texttt{FeynRules}~\cite{Alloul:2013bka}. The \texttt{Pythia8}~\cite{Sjostrand:2007gs} is used to implement parton shower, hadronization and decay of the SM hadrons. The detector effects are simulated by \texttt{Delphes}~\cite{deFavereau:2013fsa} with the default ILC configuration card. 
The leading order cross sections of the generic SM backgrounds $\mu^+ e^- \to e^+ j j +\slashed{E}_T$ and $\mu^- j j+\slashed{E}_T$ are listed in Tab.~\ref{tab:xsec}. While generating the background, the transverse momenta of the final state charged lepton and jets are required to be greater than 10 GeV and 20 GeV, respectively.
The dominant SM background processes are $\mu^+ e^- \to e^+ j j + \nu \nu \bar{\nu}$, $\mu^- j j+ \bar{\nu} \nu \bar{\nu}$, $e^- j j + \bar{\nu}$ and $\mu^+ j j+ \nu$, respectively. Their leading order cross sections are listed in Tab.~\ref{tab:xsec}. The later two processes becomes the background when the charge of the lepton is mis-identified. In our analysis, the charge mis-identification rate is taken to be 0.001~\cite{ATLAS:2019jvq}.
 \begin{table}[htbp]
\centering
\begin{tabular}{c|>{\centering\arraybackslash}p{2cm}}
\midrule [1pt]
\midrule [1pt]
Process & $\sigma$ [pb] \\
\toprule [1pt]
$\mu^+e^- \rightarrow e^+jj+\nu \nu \overline{\nu}$ & $7.97\times10^{-5}$ \\
$\mu^+e^- \rightarrow \mu^-jj+\overline{\nu} \nu \overline{\nu}$ & $8.42\times10^{-5}$ \\
$\mu^+e^- \rightarrow e^-jj+ \overline{\nu}$ & $1.33\times10^{-1}$ \\
$\mu^+e^- \rightarrow \mu^+jj+\nu$ & $1.46\times10^{-2}$ \\
\midrule [1pt]
\bottomrule [1pt]
\end{tabular}
\caption{\label{tab:xsec}The leading order production cross section of generic SM backgrounds at $\sqrt{s}=346$ in $\mu$TRISTAN experiment.  The jet in the final state should satisfy $p_T>20$ GeV and $|\eta|<4.5$, and the lepton is required have $p_T>10$ GeV and $|\eta|<2.5$.}
\end{table}
\par
\begin{figure}[htbp]
\includegraphics[width=0.4\textwidth]{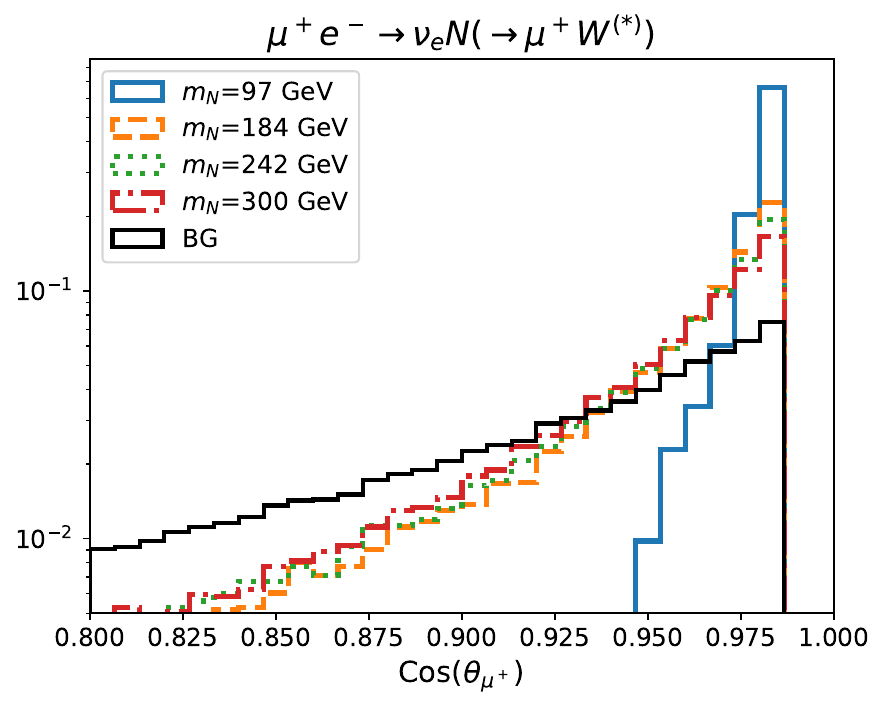}
\includegraphics[width=0.4\textwidth]{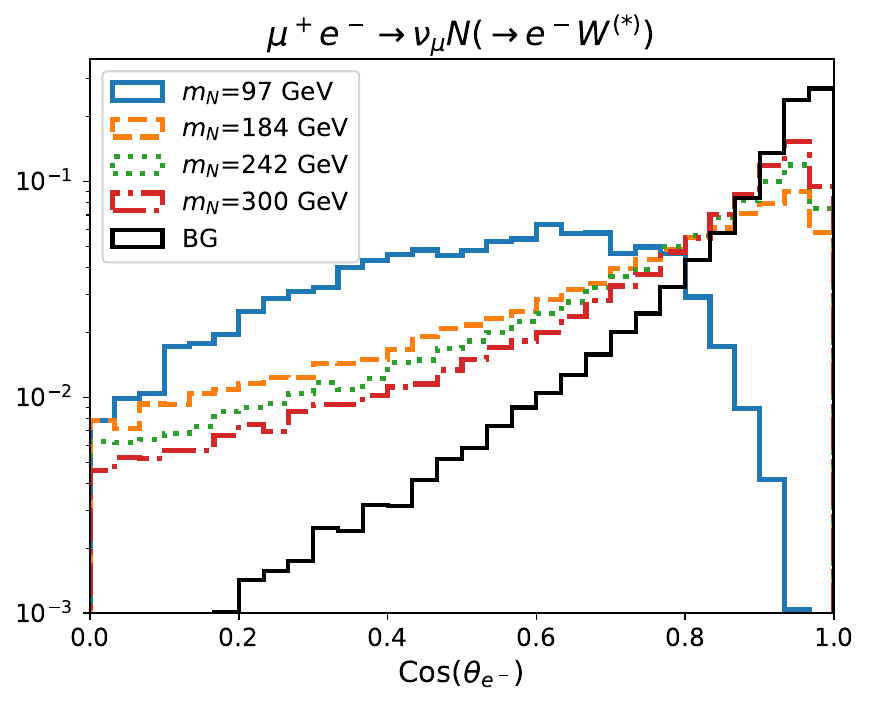}
\includegraphics[width=0.4\textwidth]{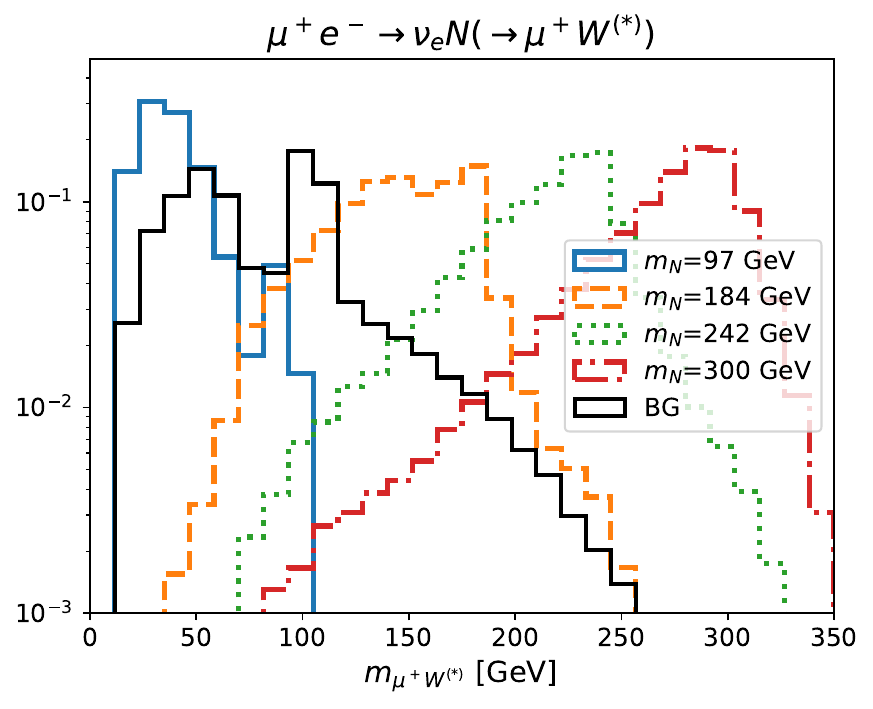}
\includegraphics[width=0.4\textwidth]{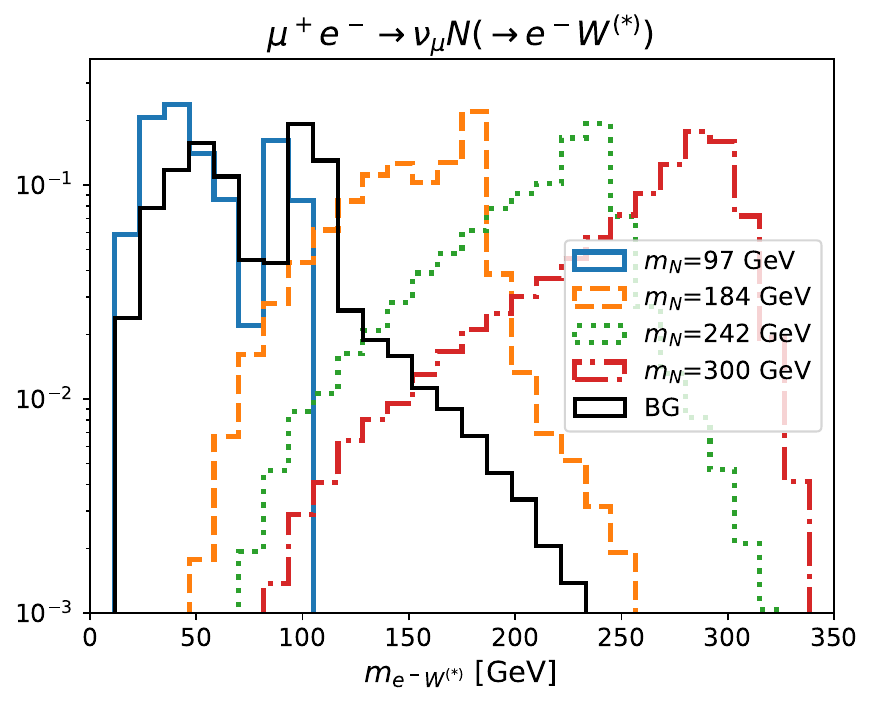}
\includegraphics[width=0.4\textwidth]{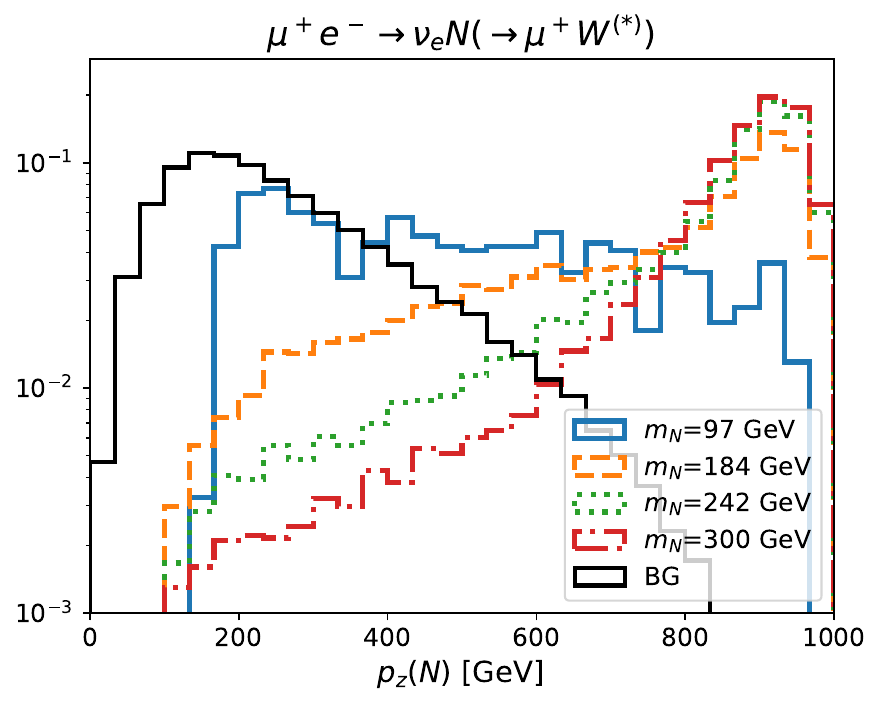}
\includegraphics[width=0.4\textwidth]{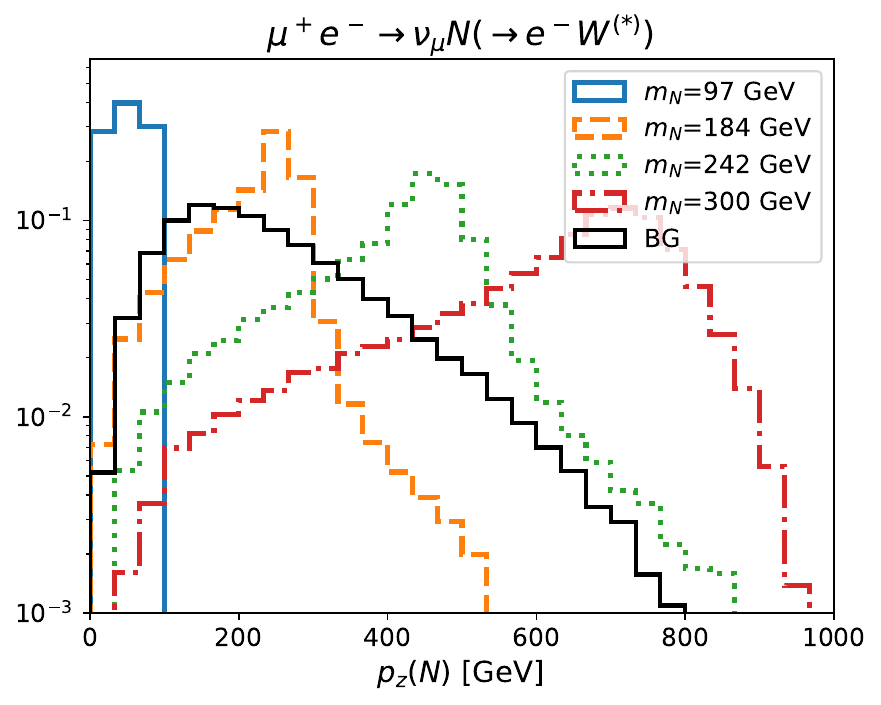}
\caption{\label{fig:vars} Feature variables that input to the BDT from top to bottom: cos$\theta$ of $\mu$/e, invariant mass and longitudinal momentum of the reconstructed heavy neutrino. Left panels: muonic flavor heavy neutrino; Right panels: electronic flavor heavy neutrino.}
\end{figure}

\begin{figure}[htbp]
\includegraphics[width=0.4\textwidth]{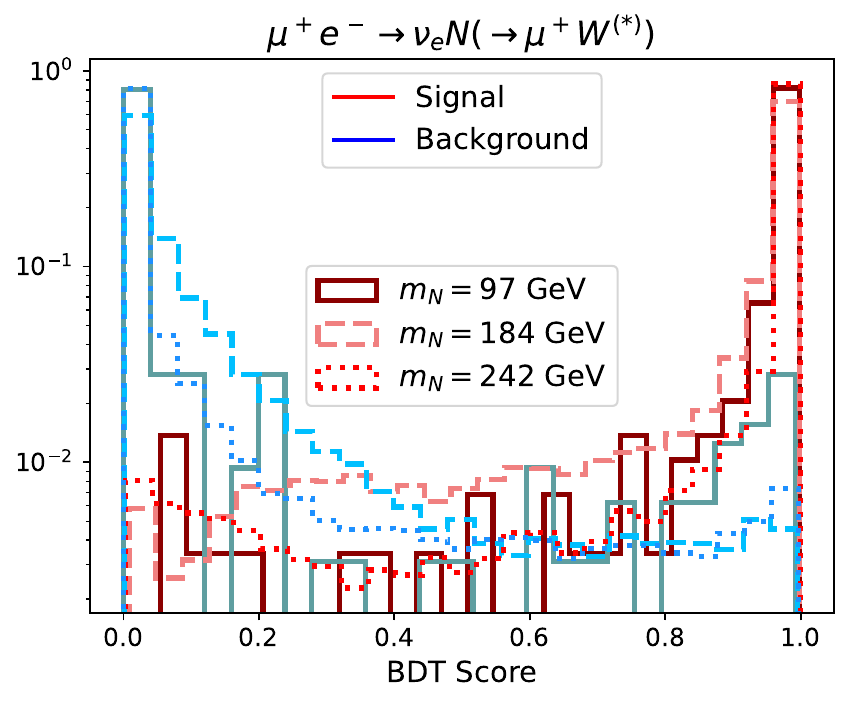}
\includegraphics[width=0.4\textwidth]{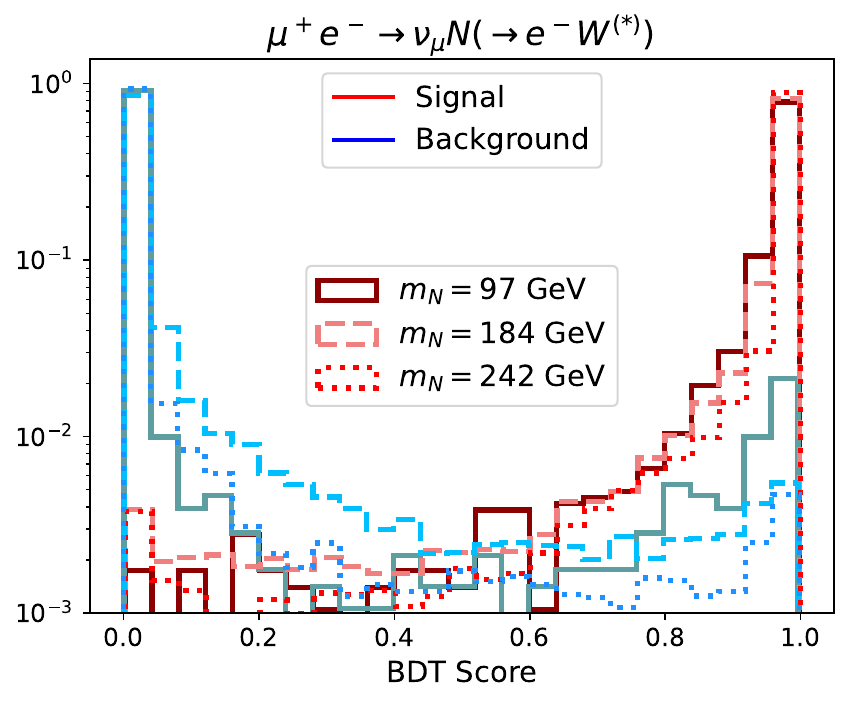}
\caption{\label{fig:bdt} The BDT score for a few benchmark signals and the corresponding backgrounds.}
\end{figure}

\begin{figure}[]
\includegraphics[width=0.497\textwidth,angle=0]{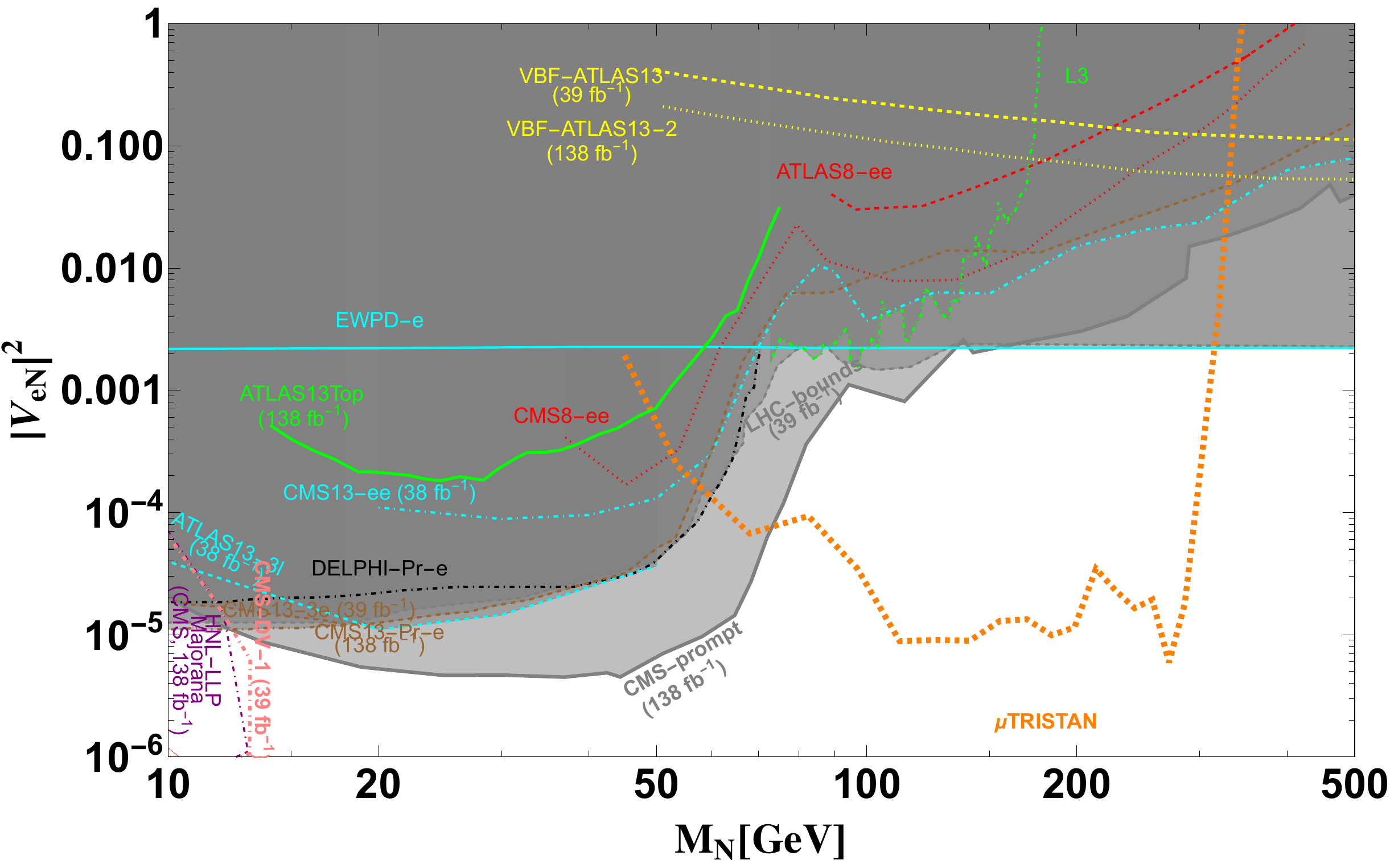}
\includegraphics[width=0.497\textwidth,angle=0]{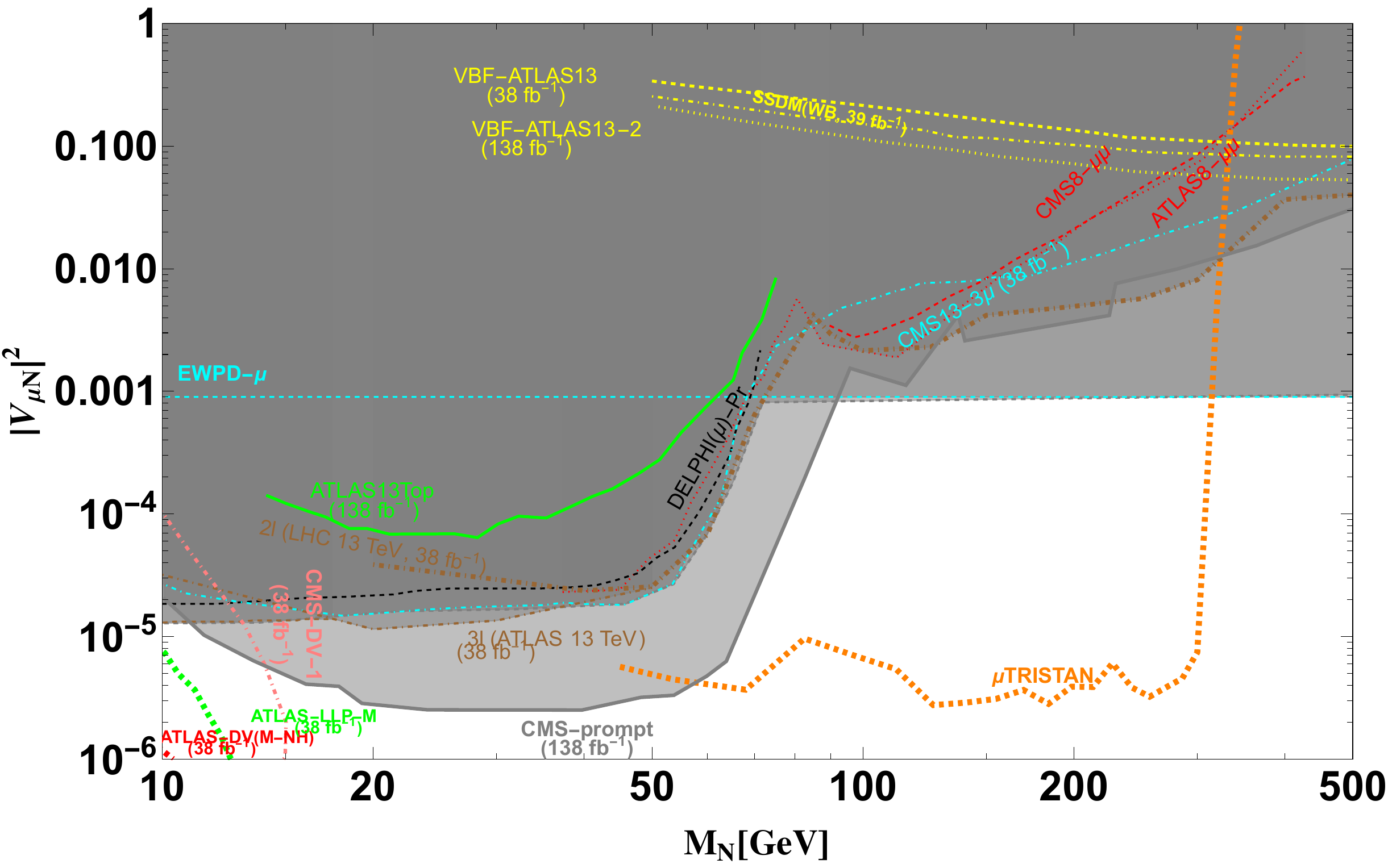}
\includegraphics[width=0.497\textwidth,angle=0]{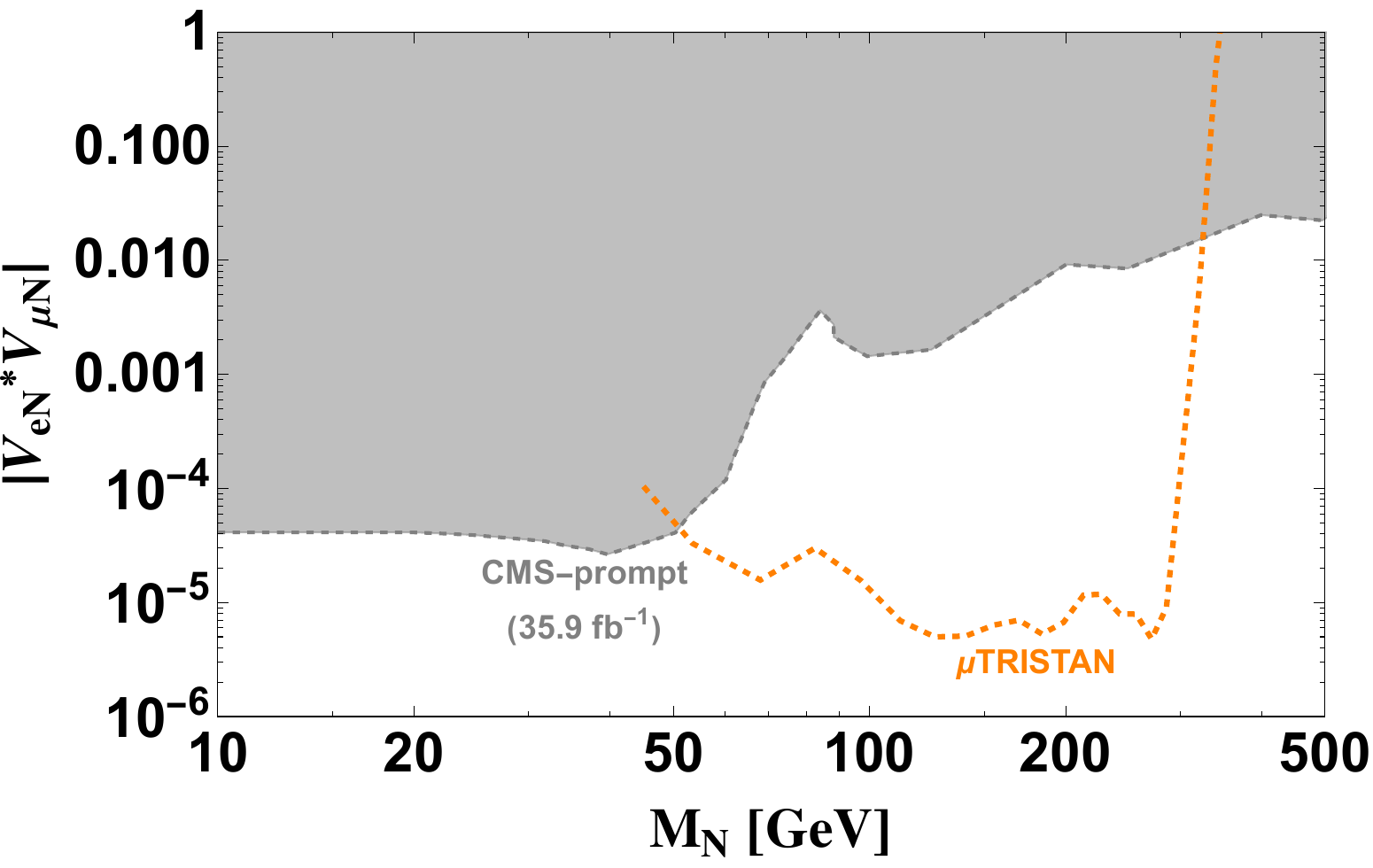}
\caption{Limits on light-heavy mixing with respect to first (left) and second (right) generation heavy neutrino masses from $\mu$TRISTAN experiment (orange dashed, thick) at $\sqrt{s}=346$ GeV and 1 ab$^{-1}$ luminosity. Shaded regions are already ruled out by different existing experimental searches (see text for other prospective limits on mixing).}
\label{fig11}
\end{figure}

 To increase the signal significance, we will adopt the {Gradient Boosted Decision Tree (GBDT)} method~\cite{Chen_2016} that takes into account several feature variables for signal and background discrimination.
Only events that fulfill the following preselection cuts will be used in the training and testing of the {GBDT}: 
(1) exactly one charged lepton with $p_T(\ell^\pm)>20$ GeV and $|\eta_\ell | < 2.5$; (2) either one or two jets with $p_T(j)>20$ GeV, $|\eta_\ell | < 4.5$, and $m_{jj} \in [70,90]$ GeV for two jets events~\footnote{The one jet final state corresponds to a boosted hadronically decaying W boson.}; (3) missing transverse momentum $\slashed{E}_T <60$ GeV. 
From the pre-selected event sample, we reconstruct the following set of observables:
\begin{align}
{p^\mu(\ell^\pm),~ \cos(\theta_{\ell^\pm}),~ p^\mu(\ell jj),~ m_{\ell jj},}
\end{align}
where $p^\mu(\ell^\pm)$ denotes the four-momentum of the isolated charged lepton, $ \cos(\theta_{\ell^\pm})$ corresponds to its polar angle in the laboratory frame, while $p^\mu(\ell jj)$ and $m_{\ell jj}$ represent respectively the four-momentum and invariant mass of the $\ell jj$ system formed by the vector sum of the lepton and dijet momenta.
The distributions for some of the feature variables are shown in Fig.~\ref{fig:vars}. 
The {GBDT} method uses a 100 tree ensemble that requires a minimum training events in each leaf node of {1} and a maximum tree depth of three, {with a learning rate of 0.1.} 
It is training on benchmark signals with heavy neutrino masses in the range of $[39,300]$ GeV with a step size 14.5 GeV. 
For each mass, half of the preselected signal ($\sim 10^4$) and background ($\sim 10^4$) events are used in the training and testing of the {GBDT} respectively. To avoid overtraining, {the sub-sample is required to be less than 0.8. }
After the training, the {GBDT} is able to assign a BDT score for each event.  The score combines all discriminating abilities of the input feature variables, which can be used for signal and background discrimination. The BDT scores for signal and background are around unit and zero, respectively. In Fig.~\ref{fig:bdt}, the BDT score for a few benchmark signals and the corresponding backgrounds are shown. The signal and background are well-separated for each case, indicating the strong discriminating power of the {GBDT}. After the application of the kinematic cuts we write down the signal and background events in Tab.~\ref{tab2} for some benchmark heavy neutrino masses $(M_{N})$ considering the decay of the heavy neutrinos in the $\ell jj$ final state.
\begin{table}[h]
\begin{center}
\begin{tabular}{|c|c|c|c|c|c|}
\hline
\multirow{3}{*}{$\sqrt{s}$ (GeV)} &\multirow{3}{*}{$M_{N_1}$ (GeV)} & \multicolumn{2}{c|}{Signal $(S)$} & %
    \multicolumn{2}{c|}{Background $(B)$} \\
\cline{3-6}
&& before cuts (pb) & after cuts (fb) & before cuts (fb) & after cuts (fb)\\
\hline
 & 45 & 24.62 & 3.7 & 0.0988 & $5.525\times 10^{-3}$ \\
 & 68 & 23.6 & 71.1 & 0.0988 & $9.375\times 10^{-4}$ \\
 346 & 97 & 23.5 & 137.3 & 0.0988 & $9.023\times 10^{-4}$\\
  & 184 & 11.5 & 430.2 & 0.0988 & $2.780\times 10^{-4}$ \\
 & 300 & 2.9 & 363.24 & 0.0988 & $3.776\times 10^{-4}$ \\
\hline
\multirow{3}{*}{$\sqrt{s}$ (GeV)} &\multirow{3}{*}{$M_{N_2}$ (GeV)} & \multicolumn{2}{c|}{Signal} & %
    \multicolumn{2}{c|}{Background} \\\cline{3-6}
&& before cuts (pb) & after cuts (fb) & before cuts (fb) & after cuts (fb)\\
    \hline
    & 45 & 24.62  & 767.4  &  0.213  & $3.681\times 10^{-4}$\\
    & 68 & 23.6 &  1242.6  & 0.213  & $6.519\times 10^{-4}$\\
346 & 97 & 23.5 &  697.6  &  0.213  & $1.089\times 10^{-3}$\\
    & 184 & 11.5 & 1675.0   & 0.213  & $8.560\times 10^{-4}$ \\
    & 300 & 2.9 & 582.1 & 0.213  & $3.473\times 10^{-4}$\\    
\hline
\hline                  
\end{tabular}
\caption{Signal normalized by light-heavy mixing squared and SM background cross sections before and after cuts for several benchmark points of the heavy neutrino masses.
The cuts here include the pre-selection cuts as well as a BDT cut that maximize the signal significance for each heavy neutrino mass.}
\label{tab2}
\end{center}
\end{table}
After the application of all the kinematic cuts we estimate a $2\sigma$ contour on the $M_{N}-|V_{e(\mu)N}|^2$ and $M_N-|V_{eN}^\ast V_{\mu N}|$ planes solving the following equation
\bea
2 = \frac{ S |V_{e(\mu) N}|^2}{\sqrt{ S|V_{e(\mu) N}|^2+ B}},
\label{signi}
\eea
where $S$ and $B$ stands for signal ( when mixing-squared is taken to be unity) and corresponding background events, respectively at 1 ab$^{-1}$ luminosity. Corresponding estimated bounds on the light-heavy mixings $|V_{eN}|^2$, $|V_{\mu N}|^2$ and $|V_{eN}^* V_{\mu N}|$ are shown in the upper left, upper right and bootom panel of Fig.~\ref{fig11}, respectively. The shaded regions in each panel are already ruled out by different existing searches. From our analysis we find that in case of the light-heavy mixing angle $|V_{eN}|^2$, prospective limits could be stronger than existing bounds from prompt heavy neutrino searches from CMS \cite{CMS:2024xdq,CMS:2024bni} and EWPD \cite{deBlas:2013gla,delAguila:2008pw,Akhmedov:2013hec} for 77 GeV $\leq M_{N} \leq 315$ GeV. We find that for 93 GeV $\leq M_{N} \leq 295$ GeV, prospective limit on $|V_{e N}|^2$ could vary in the range $4.2\times 10^{-6} \leq |V_{e N}|^2 \leq 10^{-5}$.
In addition to that, we estimate prospective limits on the mixing $|V_{\mu N}|^2$ which could provide stronger limits compared to prompt heavy neutrino searches from CMS \cite{CMS:2024xdq,CMS:2024bni} (gray solid) and EWPD (cyan solid(dashed) for electron(muon))\cite{deBlas:2013gla,delAguila:2008pw,Akhmedov:2013hec} for 61.5 GeV $\leq M_{N} \leq 321$ GeV. We find that for 88 GeV $\leq M_{N} \leq 304$ GeV, prospective limit could vary in the range $2.0 \times 10^{-6} < |V_{\mu N}|^2 \leq 10^{-5}$. 

We compare the prospective bounds with existing results from a variety of experiments represented by the gray shaded region in Fig.~\ref{fig11}. Bounds from the same sign di-lepton (SSDL) search by ATLAS and CMS experiments at $\sqrt{s}=8$ TeV LHC are shown by the red dashed and dotted lines from \cite{ATLAS:2015gtp,CMS:2015qur}. Experimental bounds from L3 detector of the LEP experiment are shown by light green dot dashed line from \cite{L3:2001zfe} in Fig.~\ref{fig11} (upper panel). Bounds from $\sqrt{s}=13$ TeV LHC using the SSDL searches from CMS \cite{CMS:2018jxx} using cyan dot-dashed and trilepton searches from ATLAS \cite{ATLAS:2019kpx} using cyan dashed and CMS \cite{CMS:2018iaf} brown dot dashed lines, respectively. Bounds obtained from the prompt (Pr) heavy neutrinos searches from the DELPHEI \cite{DELPHI:1996qcc} experiment are represented by black dot dashed and dashed lines for electron and muon flavors, respectively. Bounds on the mixing using long-lived particle (LLP) searches for Majorana type heavy neutrinos from the CMS \cite{CMS:2022fut} are shown by purple thin dot-dashed line in the upper panel of Fig.~\ref{fig11}. Displaced vertex (DV) searches in ATLAS from Majorana (M) type heavy neutrinos are taken from \cite{ATLAS:2022atq,ATLAS:2024fdw} where bounds are shown for single flavor case with normal hierarchy (NH) using red dot-dashed line, lines for muon flavor in the lower panel of Fig.~\ref{fig11}. We show corresponding CMS bounds for displaced vertex from Majorana type heavy neutrinos by pink dot dashed line \cite{CMS:2024bni,CMS:2024bni} in the upper and lower panels of Fig.~\ref{fig11}. We also estimate bounds on the light-heavy mixing from the top quark decay into heavy Majorana neutrinos shown by light green solid line in the shaded region \cite{ATLAS:2024fcs}. Limits on the light-heavy mixing due to the second generation heavy neutrino studying the vector boson fusion process from ATALS (VBF-ATLAS13, VBF-ATLAS13-2)\cite{ATLAS:2023tkz,ATLAS:2024fdw,ATLAS:2024rzi} and CMS (SSDM(W)) \cite{CMS:2022hvh} at different luminosities are shown by different yellow lines in the gray shaded region which are weaker than the bounds obtained from the EWPD. These limits could be probed in future. We show the limits on $|V_{eN}^\ast V_{\mu N}|$ in the lower panel of Fig.~\ref{fig11}. We compare our results with the bounds obtained from the 13 TeV LHC \cite{CMS:2015qur,CMS:2024bni} at 35.9 fb$^{-1}$ luminosity. We show the excluded region by LHC in gray. 

Finally we mention that limits on light-heavy mixing of neutrinos were studied from a variety of production modes and decays of the RHNs at the hadron colliders involving a variety of intial states \cite{Das:2014jxa,Das:2015toa}, considering next-to-leading order QCD corrections at the initial states \cite{Das:2016hof}, involving boosted objects like fat-jets \cite{Das:2017gke,Bhardwaj:2018lma} where we found that $\mathcal{O}(100)$ GeV RHNs could be probed within $10^{-4} \leq |V_{\ell N}|^2 \leq 10^{-3}$ at 3 ab$^{-1}$ luminosity. For RHNs lighter than $\mathcal{O}(100)$ GeV \cite{Das:2017zjc,Das:2017rsu} which could be probed from Higgs decay at LHC, the prospective limit on light-heavy mixing square could reach around $\mathcal{O}(10^{-5})$ at 3 ab $^{-1}$ luminosity. If the heavy neutrinos mass is around and below $\mathcal{O}(10)$ GeV, the bounds comes from the meson decay and could be stronger than $\mathcal{O}(10^{-5})$ \cite{Cvetic:2018elt,Cvetic:2019rms,Chun:2019nwi}. RHNs having mass $\mathcal{O}(1)$ TeV can be produced at the electron-positron colliders and considering their decay into boosted objects \cite{Banerjee:2015gca,Chakraborty:2018khw,
Das:2018usr,Mekala:2022bvr,Mekala:2022cmm} where prospective limits on light-heavy neutrino mixing square were projected around $\mathcal{O}(10^{-5})$. On the other hand, the prospective bound for the RHN mass in the range of $\mathcal{O}(10)$ GeV to $\mathcal{O}(100)$ GeV was studied in electron-positron and electron-photon colliders where the predicted bound is of the order of $\mathcal{O}(10^{-5})$ and slightly stronger than that \cite{Das:2023tna} depending on the RHN mass. We find that our obtained bound on the light-heavy mixings are comparable with most of these above mentioned prospective bounds.

\subsection{Triplet scalar search}
\begin{figure}[]
\includegraphics[width=0.4\textwidth,angle=0]{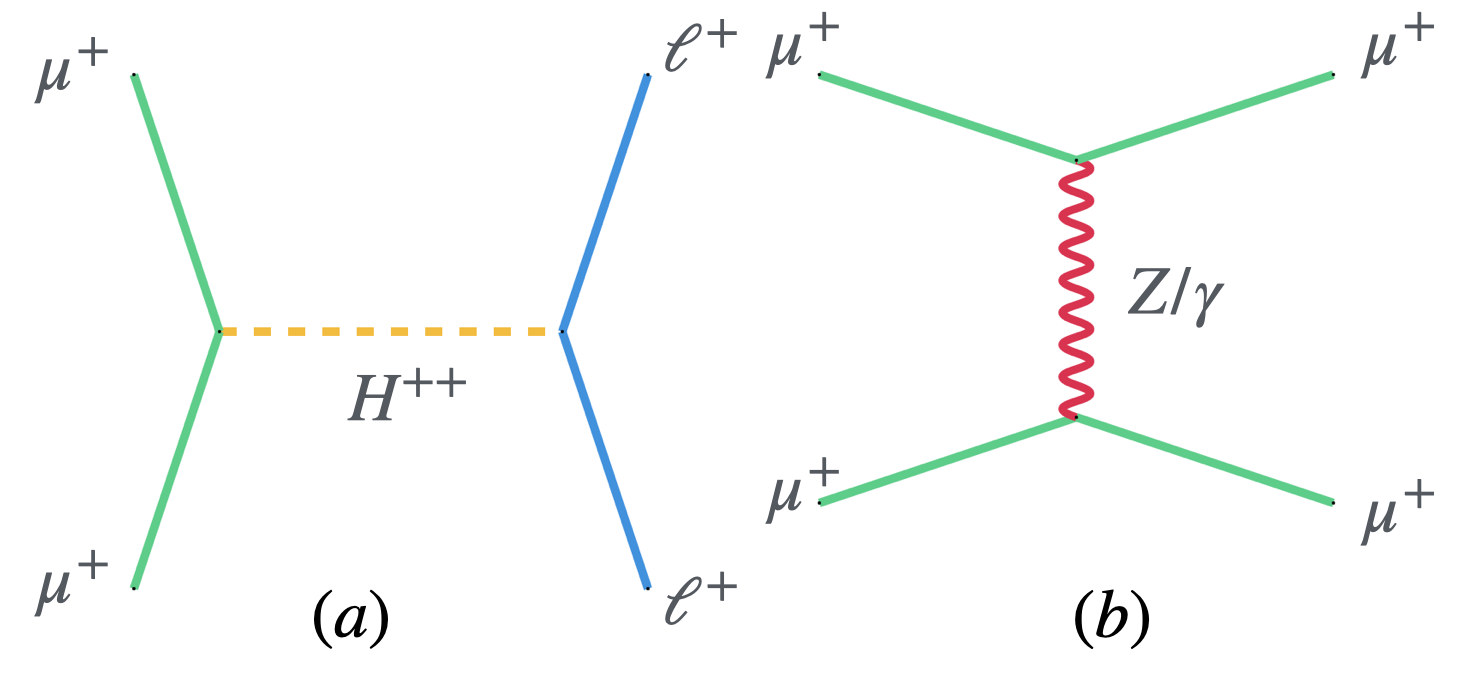}
\includegraphics[width=0.55\textwidth,angle=0]{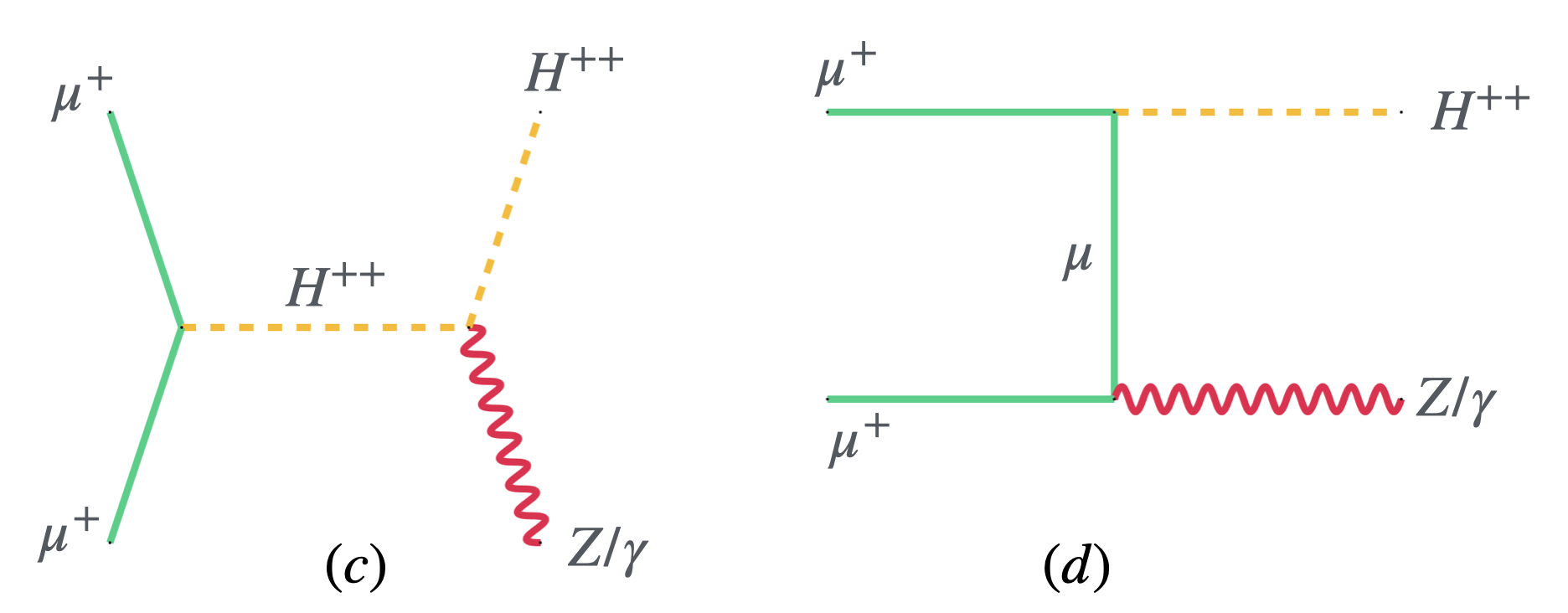}
\caption{Doubly charged scalar multiplet $(H^{++})$ mediated $\mu^+ \mu^+ \to \ell^+ \ell^+$  processes in (a) and the SM backgrounds ($\gamma, Z$) in (b) in $\mu$TRISTAN experiment within type-II seesaw framework. Feynman diagrams for $\mu^+ \mu^+ \to H^{++} Z/\gamma$ process are shown in (c) and (d) followed by the decay $H^{++} \to \ell^+ \ell^+$ which could be tested in $\mu$TRISTAN experiment at $\sqrt{s}$=2 TeV.}
\label{fig8}
\end{figure}

We consider the type-II seesaw scenario where doubly charged component of the $SU(2)$ scalar triplet could play a crucial role in $\mu^+ \mu^+$ collision of the proposed $\mu$TRISTAN collider. We focus mainly on the small value of triplet VEV $v_\Delta$ such that doubly charged Higgs exclusively decay only to the leptonic mode. The corresponding branching ratio are given by
\begin{equation}
	\text{BR}(H^{\pm\pm} \rightarrow l_i^\pm l_j^\pm) = \frac{2}{(1+\delta_{ij})}\frac{|Y_{\Delta}^{ij}|^2}{\sum_{ab} |Y_{\Delta}^{ab}|^2},
	\label{Eq:Hpp_leptonic}
\end{equation}
with $Y_\Delta$ as given in Eq.~\ref{eq:Yukawa}. The flavor structure of the Yukawa coupling $Y_\Delta$ plays a crucial role here, as it governs both the properties of neutrinos and the branching ratio of $H^{\pm\pm}$ decaying into charged leptons of various flavors. Hence, studying the branching ratios of doubly charged Higgs, one can shed some light on neutrino mass ordering. In view of this we first consider the doubly charged scalar mediated same-sign dilepton channel $\mu^+ \mu^+ \to \ell_i^+ \ell_j^+$. Note that for the final state $\ell_i^+ \ell_j^+=\mu^+\mu^+$, there would be large SM background coming from t-channel $Z/\gamma$ mediation. Corresponding Feynman diagrams are given in Fig.~\ref{fig8}. Total production cross section of $\mu^+ \mu^+ \to \ell_i^+ \ell_j^+$~($\neq \mu^+\mu^+$) signal process can be given by
\begin{align}
\sigma(\mu^+\mu^+\to \ell_i^+\ell_j^+)=\frac{|Y_\Delta^{\mu\mu}Y_\Delta^{ij}|^2}{4\pi (1+\delta_{ij})}\,\frac{s}{(s-m_{H^{\pm\pm}}^2)^2+m_{H^{\pm\pm}}^2\Gamma_{H^{\pm\pm}}^2} ,
\label{eq:mumutoll}
\end{align}
neglecting initial and final state lepton masses. Here $\sqrt{s}=2$ TeV, $\delta_{ij}$ is Kronecker delta vanishing for leptons of different kind, giving unity for same type leptons and $\Gamma_{H^{\pm \pm}}$ is the total decay width of $H^{\pm \pm}$. 
\par
\begin{figure}[]
\includegraphics[width=0.45\textwidth,angle=0]{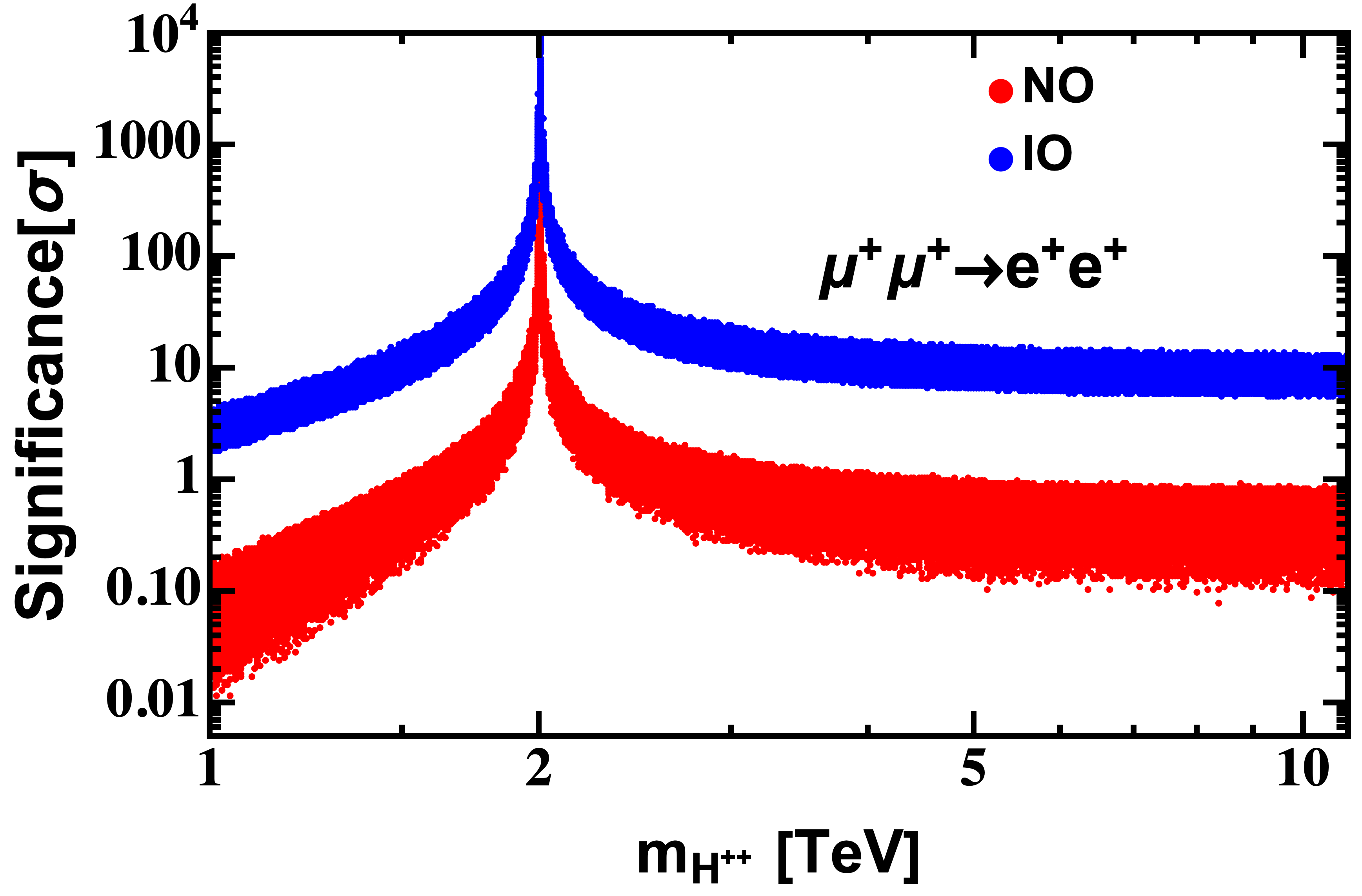}
\includegraphics[width=0.45\textwidth,angle=0]{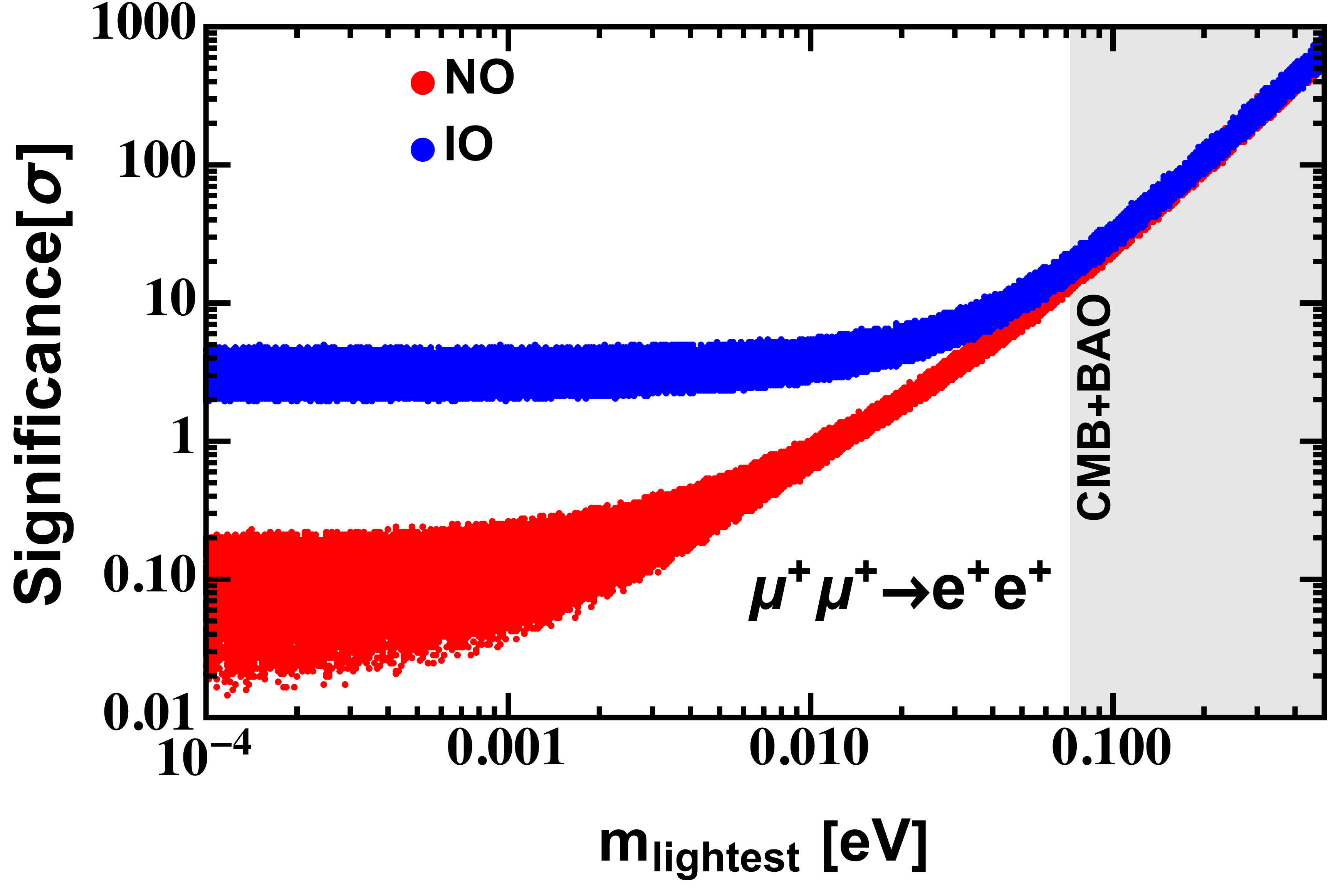}
\caption{We show the significance for $\mu^+\mu^+\to e^+ e^+$ process as function of $m_{H^{++}}$~(left panel) and lightest neutrino mass for $m_{H^{++}}=1.03$ TeV~(right panel) in $\mu$TRISTAN collider considering $\sqrt{s}=2$~TeV and luminosity as $\mathcal{L}=1\text{ ab}^{-1}$ assuming $v_{\Delta}$ and $m_{H^{++}}$ to be taken in such a way that the branching ratios of the LFV processes $\mu\to e\gamma$ and $\mu\to 3e$ are satisfied by the experimental limits $\text{BR}(\mu\to e\gamma)=4.2\times 10^{-13}$ and $\text{BR}(\mu\to 3e)=10^{-12}$ \cite{MEG:2013oxv,BaBar:2009hkt}, respectively. The neutrino oscillation parameters are varied within their allowed $3\sigma$ range \cite{deSalas:2020pgw} whereas the Majorana phases $(\phi_1,~\phi_2)$ are set to be zero. The gray shaded region in the bottom right panel is excluded from the combined analysis of CMB+BAO \cite{eBOSS:2020yzd}.}
\label{fig:ll-1}
\end{figure}

 We consider a luminosity of 1 ab$^{-1}$ to estimate number of events for the signal and background. We generate the signal and background events with transverse momentum of the leptons as $p_T^\ell > 400$ GeV and missing energy as $\slashed{E}_T<60$~GeV. As we are considering the doubly charged Higgs mass in the range $m_{H^{\pm\pm}}>1$ TeV, the transverse momentum of charged leptons coming from $H^{\pm\pm}$ decay will be very high. We find that usage of these cut does not reduce the signal that much, while background will be highly suppressed. We find that SM background is very large for $\mu^+ \mu^+$ final state as a result we did not show the results for this final state. However, for the other combinations of the signal events like $e^+ \mu^+$, $\mu^+ \tau^+$, $e^+ \tau^+$, $\tau^+ \tau^+$ and $e^+e^+$, the number of signal events are large compared to the SM backgrounds. The dominant background comes from the $WW$ scattering and becomes $\mathcal{O}(10^{-7}\text{ pb})$ after we apply the $p_T^{\ell}$ and $\slashed{E}_T$ cuts. We find that although the expected number of events can be large for all the final states, only the $e^+ e^+$ final state potentially can differentiate between the NO and IO case for neutrino mass ordering. In Fig.~\ref{fig:ll-1} we show the significance for the $\mu^+ \mu^+ \to e^+ e^+$ channel for NO and IO scenarios 
as a function of $m_{H^{++}}$~(left panel) and $m_{\rm lightest}$~(right panel), respectively. We find that the significance for NO case are different from the IO case where the significance for IO case could be at least two orders of magnitude larger than the NO case. We find that for $m_{\rm{lightest}} < 0.01$ eV, the significance for the NO and IO scenarios became nearly two orders of magnitude different. 
Therefore $\mu^+ \mu^+ \to e^+ e^+$ process mediated by $H^{++}$ can be tested in the $\mu$TRISTAN experiment to probe IO of the neutrino mass generation mechanism. In this analysis $v_{\Delta}$ and $m_{H^{++}}$ are take in such a way that the branching ratios of the LFV decay mode $\mu\to e\gamma$ and $\mu\to 3e$ are satisfied by the experimental limits $\text{BR}(\mu\to e\gamma)=4.2\times 10^{-13}$ and $\text{BR}(\mu\to 3e)=10^{-12}$ \cite{MEG:2013oxv,BaBar:2009hkt}, respectively. The neutrino oscillation parameters are varied within their allowed $3\sigma$ range \cite{deSalas:2020pgw} whereas the Majorana phases $(\phi_1,~\phi_2)$ are set to be zero. The gray shaded region in Fig.~\ref{fig:ll-1} is excluded from the combined analysis of CMB+BAO \cite{eBOSS:2020yzd}. 
\ignore{We found that the generic SM background is negligibly small, therefore depending on $m_{H^{++}}$ and $m_{\rm{lightest}}$ IO scenarios could be observed at the $\mu$TRISTAN experiment with at least 3$\sigma$ significance or more considering $e^+ e^+$ final state with luminosity varying between 100 fb$^{-1}$ to 1 ab$^{-1}$. }
\par We study another production mode of $H^{++}$ following $\mu^+ \mu^+ \to H^{++} \gamma/ Z$ and the corresponding Feynman diagrams are given in Fig.~\ref{fig8}. We fist consider the scenario  $\mu^+\mu^+\to H^{++}\gamma\to e^+ e^+\gamma$ where doubly charged Higgs decays to $e^+e^+$ final state. The dominant SM background for this channel comes from the $W$-boson fusion channel such as $\mu^+\mu^+\to W^+ W^+ + \bar{\nu}\bar{\nu}$, following the leptonic decay of $W$ boson and photons radiated from any of the charged leptons. In this analysis we apply the following cuts: $p_T^\gamma>50$~GeV,  $p_T^e>400$~GeV and $\slashed{E}_T<60$~GeV, respectively. After cuts, signal events become  almost half of the events before cut, whereas the SM background reduces to $\sim 10^{-9}$ pb after the application of kinematic cuts from the cross section $\sim 10^{-3}$ pb obtained before  application of kinematic cuts. These SM backgrounds could be large for $\mu^+\mu^+$ mode due to the following $W$-boson fusion channel such as $\mu^+\mu^+\to \mu^+\mu^+ W^+ W^-$ as well as due to the $t-$ channel $\gamma/Z$ mediated SM process $\mu^+\mu^+\to \mu^+\mu^+$. Therefore we do not consider $H^{++}\to \mu^+ \mu^+$ mode. 
\par
\begin{figure}[]
\includegraphics[width=0.45\textwidth,angle=0]{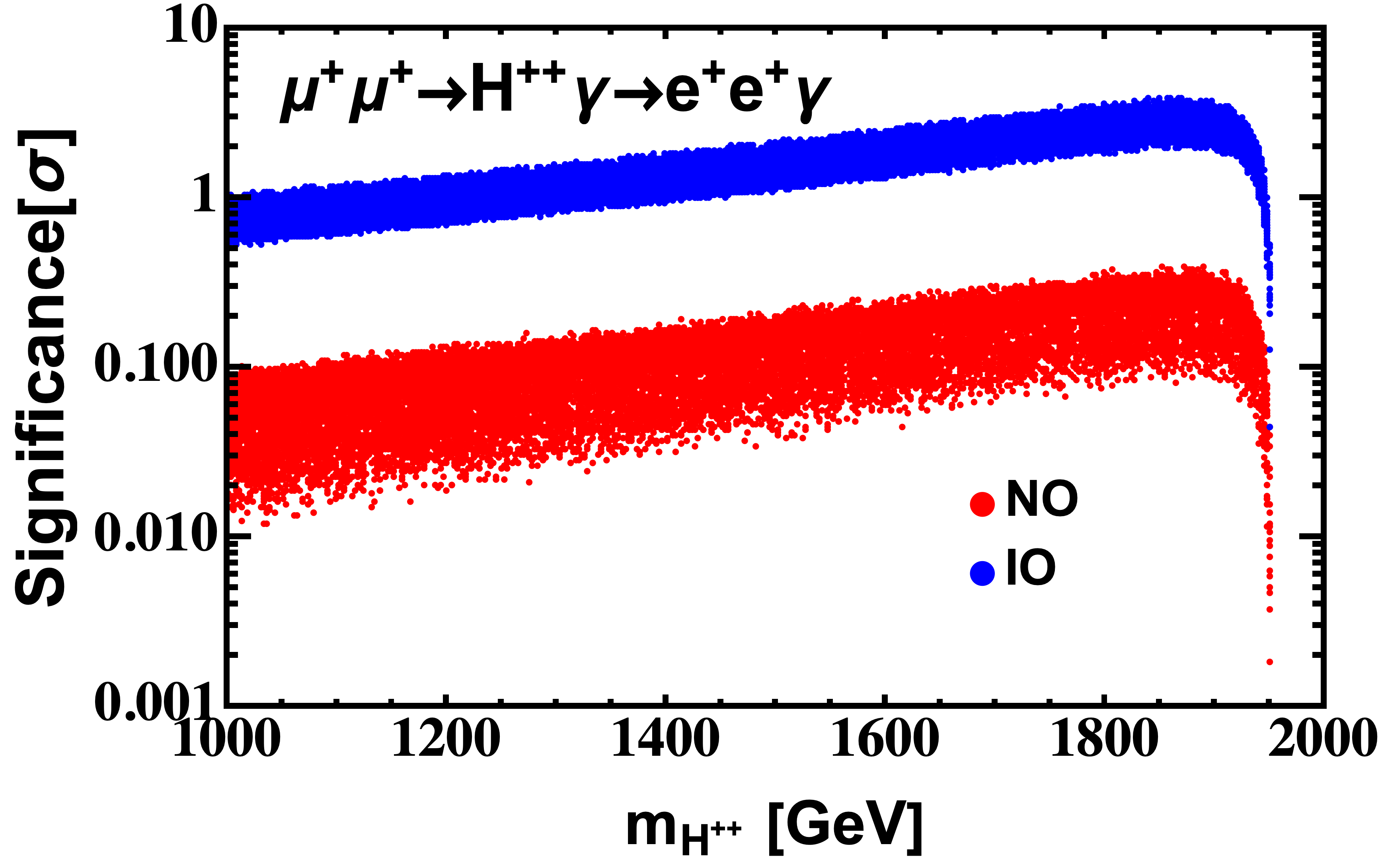}
\includegraphics[width=0.45\textwidth,angle=0]{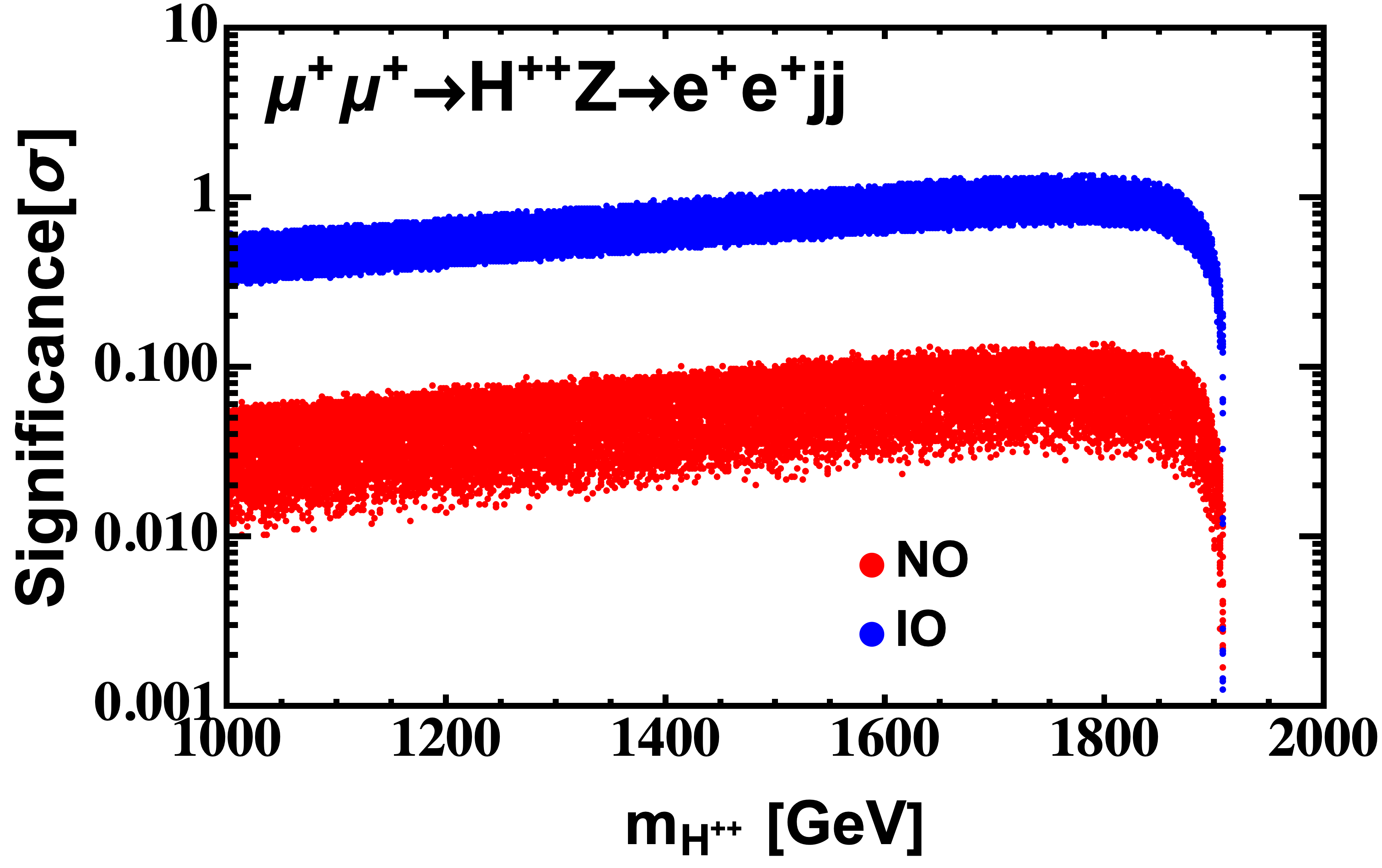}
\includegraphics[width=0.45\textwidth,angle=0]{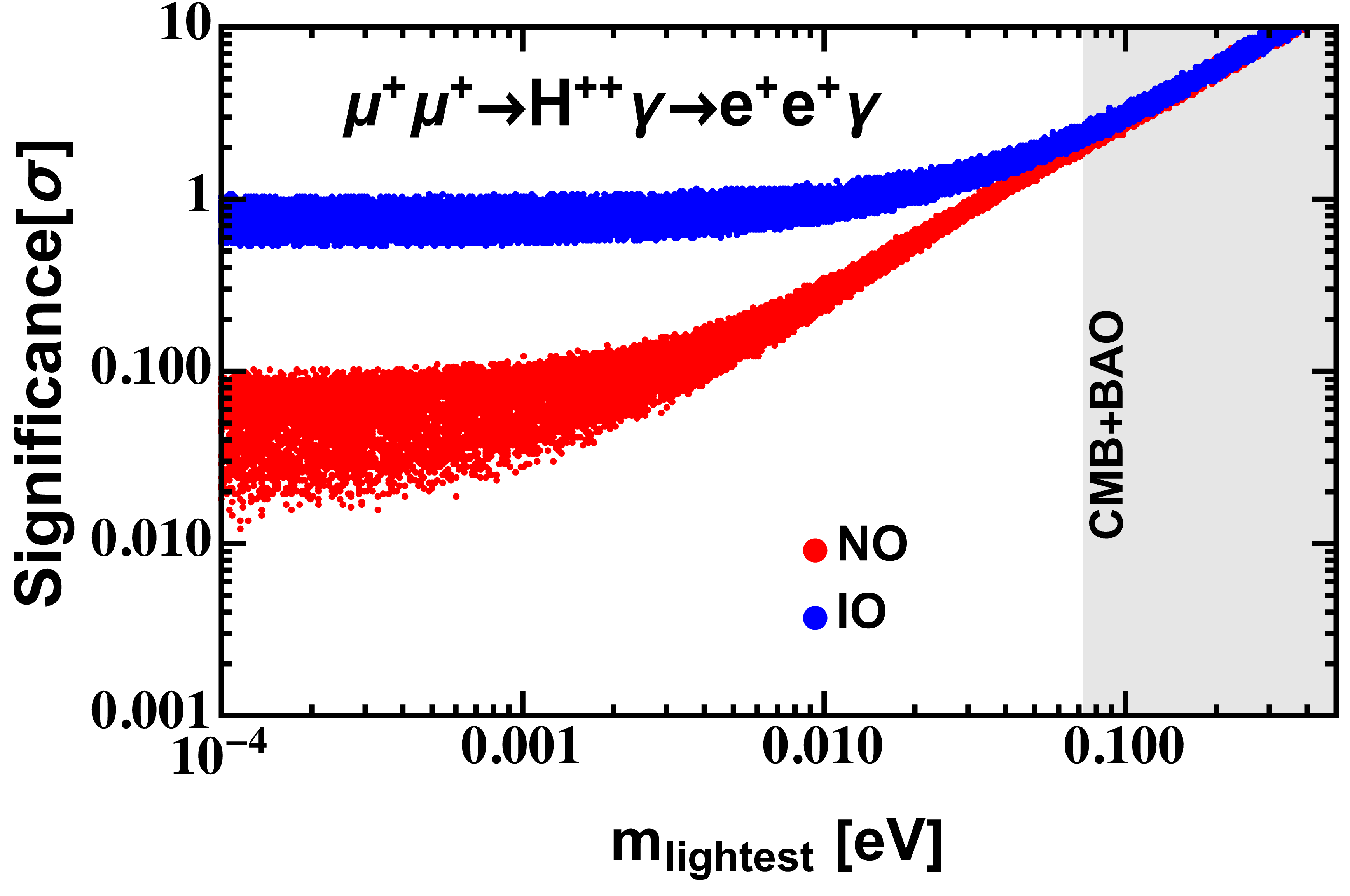}
\includegraphics[width=0.45\textwidth,angle=0]{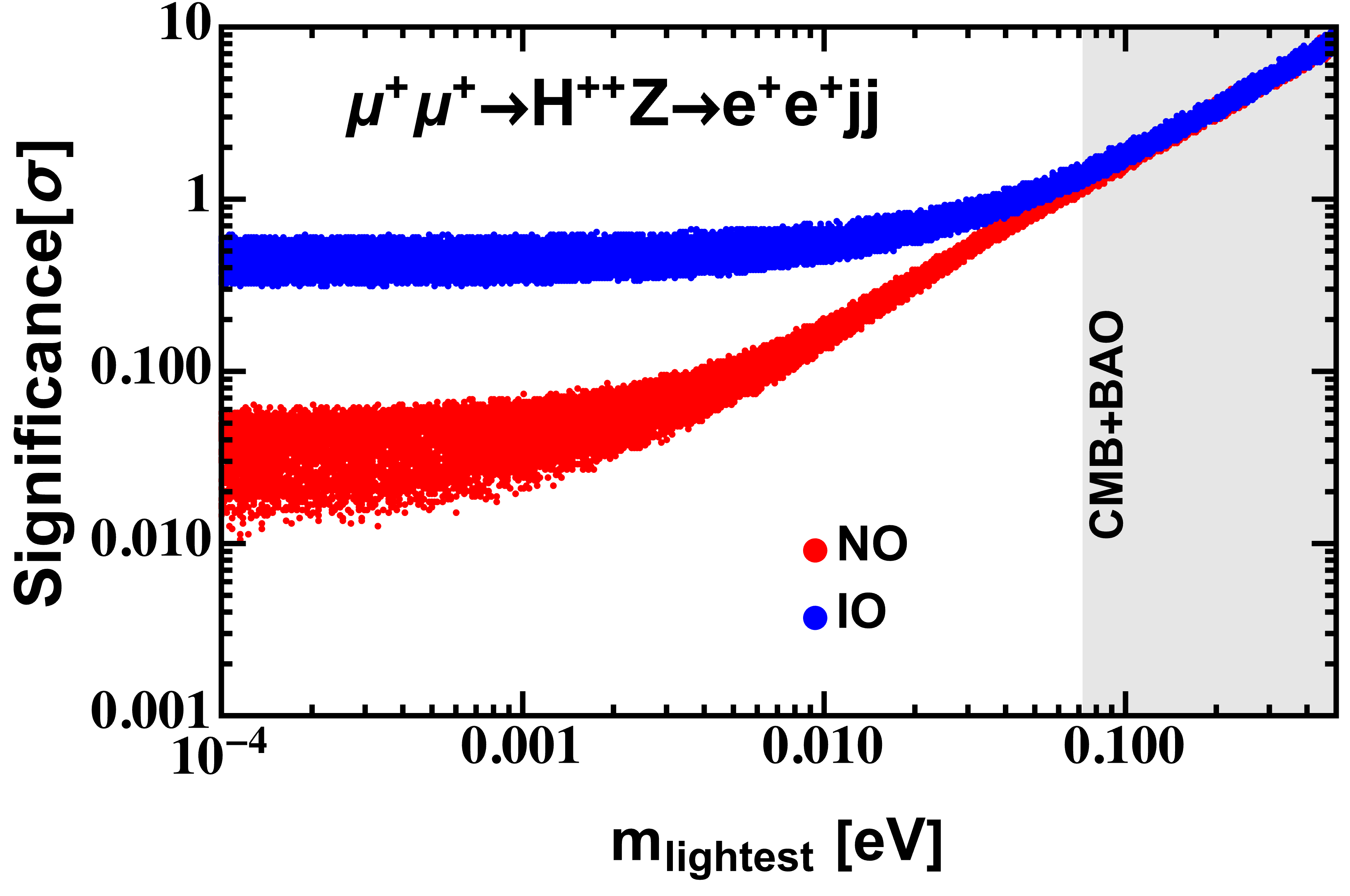}
\caption{Significance of the $\mu^+\mu^+\to e^+ e^+\gamma$~(upper left panel) and $e^+e^+jj$~(upper right panel) processes after cut with $\sqrt{s}=2$~TeV and $m_{\rm lightest}=0$ in the $\mu$TRISTAN experiment. Fixing $m_{H^{++}}=1.03$ TeV distinction between NO and IO cases are shown (bottom panel) as a function of $m_{\rm{lightest}}$. We assume $v_{\Delta}$ and $m_{H^{++}}$ to be taken in such a way that the branching ratios of the LFV processes $\mu\to e\gamma$ and $\mu\to 3e$ are satisfied by the experimental limits $\text{BR}(\mu\to e\gamma)=4.2\times 10^{-13}$ and $\text{BR}(\mu\to 3e)=10^{-12}$ \cite{MEG:2013oxv,BaBar:2009hkt}, respectively. The neutrino oscillation parameters are varied within their allowed $3\sigma$ range \cite{deSalas:2020pgw} whereas the Majorana phases $(\phi_1,~\phi_2)$ are set to be zero. The gray shaded region in the bottom right panel is excluded from the combined analysis of CMB+BAO \cite{eBOSS:2020yzd}.}
\label{fig:Hpp}
\end{figure}

 Next we consider the process $\mu^+\mu^+\to H^{++} Z\to e^+ e^+ jj$, see Fig.~\ref{fig8}. We are considering hadronic decays of $Z$ boson due to larger branching ratio over leptonic mode and smaller SM background. To analyze the signal and SM background events we apply kinematic cuts as $p_T^e>400$~GeV, $p_T^j>20$~GeV and $\slashed{E}_T<60$~GeV. The dominant SM background for this channel comes from the following SM processes: $W^+W^++\bar{\nu}\bar{\nu}$ and $W^+W^+ (Z/h) +\bar{\nu}\bar{\nu}$. Signal events after the cuts is almost $84\%$ of the events before cuts and SM background reduces to $10^{-10}$ pb. The upper and bottom panel of Fig.~\ref{fig:Hpp} shows the significance as a function of $m_{H^{\pm\pm}}$ and $m_{\rm lightest}$, respectively. We see that as long as lightest neutrino mass is in the range $m_{\rm lightest}\leq 0.01$~eV, depending on the doubly charged Higgs mass, the significance for both the processes $e^+ e^+ \gamma$ and $e^+ e^+ jj$ clearly differs for NO case compare to the IO case.
\ignore{In the lower panel of Fig.~\ref{fig:Hpp} we show difference between the cross sections of the same final states as a function of $m_{\rm{lightest}}$ for the NO and IO cases and the difference between these two orderings of the neutrino mass comes prominent for $m_{\rm{lightest}} \leq 0.01$ eV. Depending on luminosity between 100 fb$^{-1}$ to 1 ab$^{-1}$ the significance of finding different neutrino mass hierarchies could be at least $3\sigma$ or more.}
\par
\begin{figure}[]
\includegraphics[width=0.32\textwidth,angle=0]{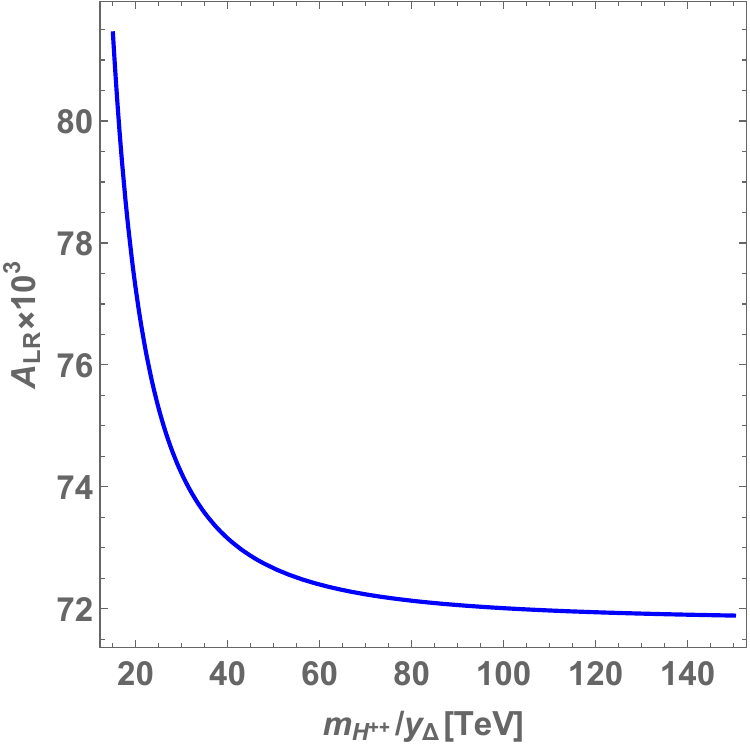}
\includegraphics[width=0.31\textwidth,angle=0]{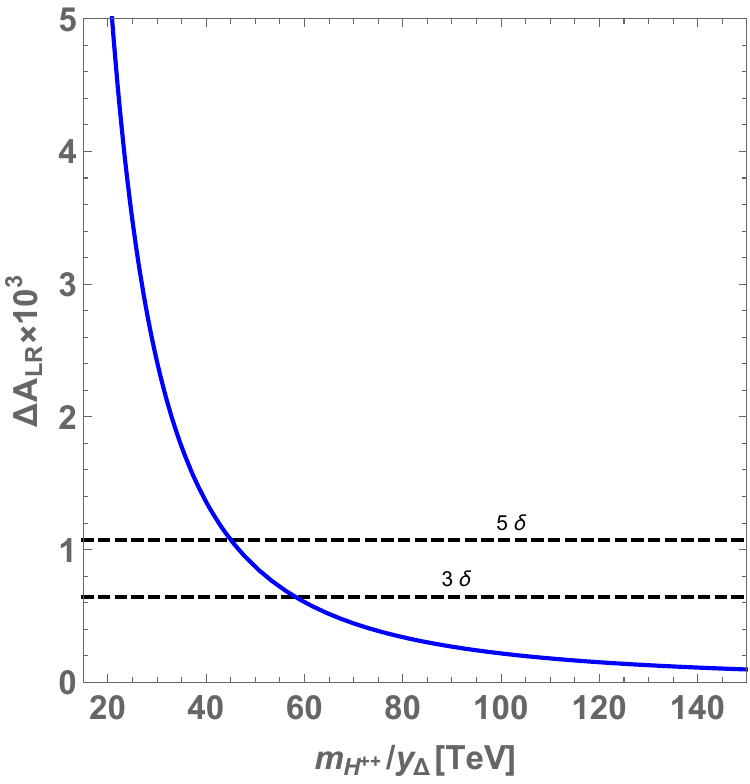}
\includegraphics[width=0.332\textwidth,angle=0]{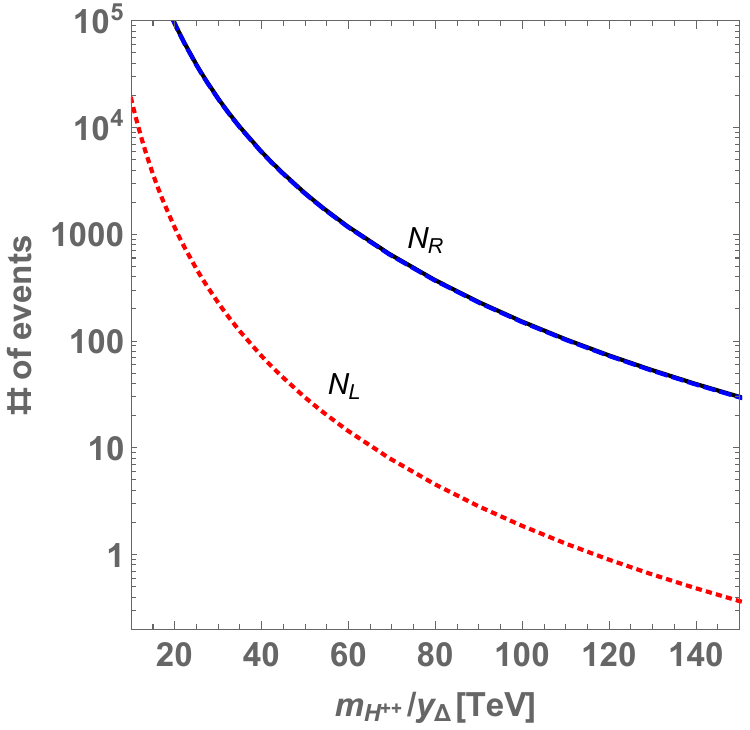}
\caption{Left right asymmetry (left most panel) and the deviation (middle panel) from the SM at 3$\sigma$ (lower black dashed line) and 5$\sigma$ (upper black dashed line) significance as a function of $m_{H^{++}}/Y_{\Delta}$ for $\mu^+\mu^+ \to \mu^+ \mu^+$ process in $\mu$TRISTAN experiment at $\sqrt{s}=2$ TeV at 1 ab$^{-1}$ luminosity. In the right most panel we show total number of events for $e^+ e^+$ and $\tau^+ \tau^+$ final states for left $(N_L)$ and right $(N_R)$ polarizations represented by red dotted and blue-black solid lines.}
\label{fig9}
\end{figure}

 Finally we study the left-right asymmetry for the process $\mu^+\mu^+\to \ell^+\ell^+$ using the polarized cross-section at the $\mu$TRISTAN experiment considering $\sqrt{s}=2$ TeV and 1 $\text{ab}^{-1}$ luminosity. Corresponding Feynman diagrams are given in Fig.~\ref{fig8}. We will mostly follow the Ref.~\cite{Nomura:2017abh} for the methodology. The relevant scattering process is given by 
\begin{equation}
  \mu^+(k_1,\sigma_1) + \mu^+(k_2,\sigma_2) \to \ell^+   (k_3,\sigma_3) + \ell^+(k_4,\sigma_4) , 
 \end{equation} 
where $\sigma_i~(k_i)$ indicate helicity~(momenta) and $\ell = \{e, \mu, \tau \}$. If the doubly charged Higgs mass is very high, the effective Lagrangian responsible for the doubly charged Higgs mediated process $\mu^+\mu^+\to\ell^+\ell^+$ can be described as
\ignore{In addition to the SM interactions we take into account doubly charged scalar exchanging interaction described by effective operator}
\begin{equation}
\mathcal{L}_{\rm eff} = \frac{(Y_\Delta^\dagger Y_\Delta)_{\mu \ell}}{2 m_{H^{++}}^2} (\bar{\mu} \gamma^\mu P_L \mu) (\bar{\ell} \gamma_\mu P_L \ell).
\end{equation}
When the final state is $\ell^+\ell^+=\mu^+\mu^+$, there will be additional contribution from SM interaction, see Fig.~\ref{fig8}. The squared helicity amplitudes $|\mathcal{ M}_{\{\sigma_i\}}|^2=|\mathcal{M}(\sigma_1\sigma_2\sigma_3\sigma_4)|^2$ for the $\mu^+\mu^+ \to \mu^+ \mu^+$ process are calculated as
\begin{align}
  & |{\cal M}(+-+-)|^2 =  |{\cal M}(-+-+)|^2 = s^2 \left(1+\cos\theta\right)^2  \left[ \frac{e^2}{t} + \frac{g_L g_R}{t_Z} \right]^2, \\ 
  & |{\cal M}(+--+)|^2 = |{\cal M}(-++-)|^2 = s^2 \left(1-\cos\theta\right)^2  \left[ \frac{e^2}{u} + \frac{g_L g_R}{u_Z} \right]^2, \\ 
  & |{\cal M}(++++)|^2 = 4s^2   \left[ e^2 \left( \frac{1}{t} +\frac{1}{u} \right) + g_R^2 \left( \frac{1}{t_Z} +\frac{1}{u_Z} \right)   \right]^2, \\ 
  & |{\cal M}(----)|^2 = 4s^2   \left[ e^2 \left( \frac{1}{t} +\frac{1}{u} \right) + g_L^2 \left( \frac{1}{t_Z} +\frac{1}{u_Z} \right) -  \frac{(Y_\Delta^\dagger Y_\Delta)_{\mu \mu}}{m_{H^{++}}^2}   \right]^2, 
\end{align}
where $g_L =e (-1/2 + s_W^2)/(s_Wc_W)$, $g_R = e s_W/c_W$, $t=-s(1-\cos\theta)/2$, $u=-s(1+\cos\theta)/2$,  $u_Z=u-m_Z^2+im_Z\Gamma_Z$, $t_Z=t-m_Z^2+im_Z\Gamma_Z$ and $\cos\theta$ is the scattering polar angle. 
\ignore{$s=(k_1+k_2)^2=(k_3+k_4)^2$,
$t=(k_1-k_3)^2=(k_2-k_4)^2=-s(1-\cos\theta)/2$, $u=(k_1-k_4)^2=(k_2-k_3)^2=-s(1+\cos\theta)/2$, 
$u_Z=u-m_Z^2+im_Z\Gamma_Z$, $t_Z=t-m_Z^2+im_Z\Gamma_Z$, 
and $\cos\theta$ is the scattering polar angle. }
For $\mu^+ \mu^+ \to e^+e^+(\tau^+ \tau^+)$ process, we do not have the SM contribution and the amplitude is only 
\begin{equation}
   |{\cal M}'(----)|^2 = 4s^2   \left[\frac{(Y_\Delta^\dagger Y_\Delta)_{\mu e(\tau)}}{m_{H^{++}}^2}    \right]^2. 
\end{equation}
Here we write differential cross-section for purely-polarized initial-state as
\begin{align}
 \frac{d\sigma_{\sigma_1\sigma_2}}{d\cos\theta} = \frac{1}{32\pi s}
 \sum_{\sigma_3,\sigma_4} \left|{\cal M}_{\{\sigma_i\}}\right|^2.
\end{align}
For realistic case, we consider partially-polarized initial-state with the degree of
polarization $P_{\mu^+}$ for the $\mu^+$ beams, and the polarized cross section is given as
\begin{align}
  \frac{d\sigma(P_{\mu^+},P_{\mu^+})}{d\cos\theta} & =
 \frac{1+P_{\mu^+}}{2} \frac{1+P_{\mu^+}}{2}\frac{d\sigma_{++}}{d\cos\theta}
 + \frac{1+P_{\mu^+}}{2}
 \frac{1-P_{\mu^+}}{2}\frac{d\sigma_{+-}}{d\cos\theta}\nonumber \\
 & + \frac{1-P_{\mu^+}}{2}
 \frac{1+P_{\mu^+}}{2}\frac{d\sigma_{-+}}{d\cos\theta} 
 + \frac{1-P_{\mu^+}}{2}
 \frac{1-P_{\mu^+}}{2}\frac{d\sigma_{--}}{d\cos\theta}.
\end{align}
For $\mu$TRISTAN, we consider the following right-handed and left-handed cases: 
\begin{align}
 & \frac{d\sigma_R}{d\cos\theta} = 
 \frac{d\sigma(P_{\mu^+}= -0.8, P_{\mu^+} = -0.8)}{d\cos\theta}, \\
 & \frac{d\sigma_L}{d\cos\theta} = 
 \frac{d\sigma(P_{\mu^+}= 0.8,P_{\mu^+}=0.8)}{d\cos\theta}.
\end{align}
To study the sensitivity to the doubly-charged scalar effects on $\mu^+ \mu^+ \to \mu^+ \mu^+$ scattering process, we consider the left-right (LR) asymmetry defined by
\begin{align}
 A_{LR} = \frac{N_R-N_L}{N_R+N_L}, 
\end{align}
where 
\begin{equation}
  N_{L(R)} =  \mathcal{L} \cdot \int_{-0.95}^{0.95}
 d\cos\theta\frac{d\sigma_{L(R)}}{d\cos\theta}, 
\end{equation}
where $\mathcal{L}=1$ ab$^{-1}$ is the luminosity we consider in this analysis. To estimate total cross sections we integrate over the range $-0.95\leq \cos\theta\leq 0.95$ of the scattering angle. 
Note that the sensitivity to new physics decreases when we integrate over the range $|\cos \theta| \leq 1$ since photon exchanging contribution is large around $|\cos \theta| \simeq 1$. Here we adopt the range taking from the previous study in ref.~\cite{Nomura:2017abh}. We also consider a statistical error of the asymmetry given by
\begin{align}
 \delta A_{LR} = \sqrt{\frac{1-A^2_{LR}}{N_R+N_L}}.
\end{align}
The sensitivity to the new physics scenario is estimated at some confidence level $(x~\sigma)$ by requiring
\begin{align}
 \left|A_{LR}({\rm SM}+\delta^{++})-A_{LR}({\rm SM})\right|\ge x\times \delta  A_{LR}({\rm SM}),
\end{align}
where $A_{LR}({\rm SM})[\delta A_{LR }({\rm SM}) ]$ can be obtained by considering the limit $m_{H^{++}} \to \infty$. For $\mu^+ \mu^+ \to e^+e^+(\tau^+\tau^+)$ process, we simply consider number of event $N_{L,R}$ since there is no SM contribution. The corresponding results are shown in the left and middle panels of Fig.~\ref{fig9} where we estimate LR asymmetry $(A_{LR})$ for the $\mu^+\mu^+ \to \mu^+ \mu^+$ process and its corresponding deviation $(\Delta A_{LR})$ from the SM as a function of $m_{H^{++}}/Y_{\Delta}$. We find that the deviation could be probed by the $\mu$TRISTAN collider at 3$\sigma$ (upper black dashed line) and 5$\sigma$ (lowe black dashed line) levels with $\sqrt{s}=2$ TeV and $\mathcal{L}=1$ ab$^{-1}$ for $m_{H^{++}}/Y_{\Delta} > 45$ TeV. 
 Note that we here assume no systematic uncertainty. In fact we find $\Delta A_{LR}/A_{LR} \simeq 0.003 $ at $m_{H^{++}}/y_{\Delta} = 50$ TeV, and roughly speaking, systematic uncertainty would need to be less than 0.3\% level to obtain sensitivity to new physics at the point. It can be feasible as the systematic uncertainty is considered to be 0.2$\%$ level for leptonic signals at the ILC~\cite{ILC:2013jhg}.
In the right most panel of Fig.~\ref{fig9} we show the number of events for left $(N_L)$ and right $(N_R)$ polarizations represented by red dotted and blue-black solid lines obtained from $e^+ e^+$ and $\tau^+ \tau^+$ final states as a function of $m_{H^{++}}/Y_{\Delta}$
where there is no SM background. We find that $N_R > N_L$ by two orders of magnitude through out the range of $m_{H^{++}}/Y_{\Delta}$ under consideration. In $\mu$TRISTAN experiment we could probe a high scale of triplet mass though sizable left-right asymmetry for different same sign dilepton mode and deviation of left-right asymmetry from the SM scenario for $e^+ e^+$ final state. 

\subsection{Triplet fermion search}
\begin{figure}[h]
\centering
\includegraphics[width=0.6\textwidth,angle=0]{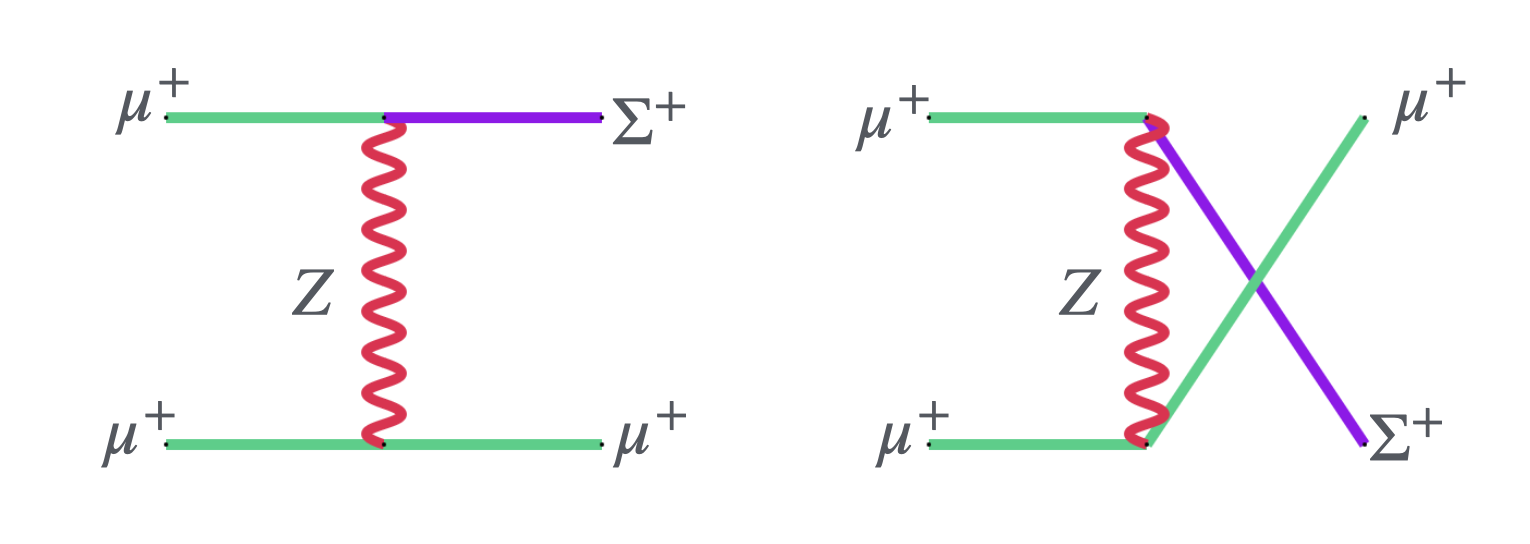}
\caption{Production of $\Sigma^+$ in $t-$channel (left) and $u-$channel (right) process in association with $\mu^+$ in $\mu$TRISTAN followed by the decay $\Sigma^+ \to \mu^+ Z$ and $Z\to j j$ in a $\mu^+\mu^+ j j$ final state.}
\label{fig10}
\end{figure}

At $\mu^+ \mu^+$ collider of the $\mu$TRISTAN experiment, $\Sigma^+$ production can be produced from the $t-$ and $u-$ channel processes following $\mu^+ \mu^+ \to \Sigma^+ \mu^+$ mediated by $Z$ boson  as shown in Fig.~\ref{fig10}. 
The differential scattering cross section of $\mu^+ \mu^+ \to \Sigma^+ \mu^+$ process including the effect of interference can be given by
\bea
 \frac{d\sigma(\mu^+ \mu^+ \to \Sigma^+ \mu^+)}{d\cos\theta} &=&(3.89\times 10^8 \times 10^3~{\rm fb}) \times \frac{s-M_\Sigma^2}{32 \pi s^2} (|M_1|^2 + 2 {\rm Re}(M_1 M_2^*) + |M_2|^2), \\
 |M_{1,2}|^2 &=& \frac{g^4}{2 c_W^4} \frac{\left(-\frac12 + s_W^2 \right)^2}{\left(\frac12 (s-M_\Sigma^2)(1 \mp \cos \theta+m_Z^2) + \Gamma_Z^2 m_Z^2 \right)}  \\
& & \Biggl[ \left(4 s_W^4 - 2 s_W^2 +\frac12 \right) \left(\frac14 s (s-M^2_{\Sigma} + \frac14(s-M_\Sigma^2)(1 \pm \cos \theta) \right)(\frac12 s -\frac14 (s-M_\Sigma^2)(1 \mp \cos\theta))  \nonumber \\
&& + \left(\frac12 - 2 s_W^2 \right)\left(\frac14 s (s- M_\Sigma^2)-\frac14(s-M^2_\Sigma)(1 \mp \cos\theta)\left(\frac12 s - \frac14 s(s-M^2_\Sigma)(1 \mp \cos\theta)\right) \right) \Biggr], \nonumber \\
2 {\rm Re}(M_1 M_2^*) &=& - \frac{g^4}{16 c_W^4} \frac{s}{2} \left(-\frac12 + s_W^2 \right)^2 (4 s_W^4 + 4 s_W^2 +1) (s - M_\Sigma^2) \nonumber \\
&& \frac{\left(\frac12(s-M^2_\Sigma)(1-\cos\theta)+m_Z^2\right)\left(\frac12(s-M^2_\Sigma)(1+\cos\theta)+m_Z^2\right) + \Gamma^2_Z m^2_Z}{\left( \left(\frac12(s-M^2_\Sigma)(1-\cos\theta) + m_Z^2\right)^2 + \Gamma_Z^2 m_Z^2  \right)\left(\left( \frac12 (s-M^2_\Sigma)(1+\cos\theta)+m_Z^2\right)^2+\Gamma_Z^2 m_Z^2 \right)}.
\eea
\begin{figure}[t]
\centering
\includegraphics[width=0.5\textwidth,angle=0]{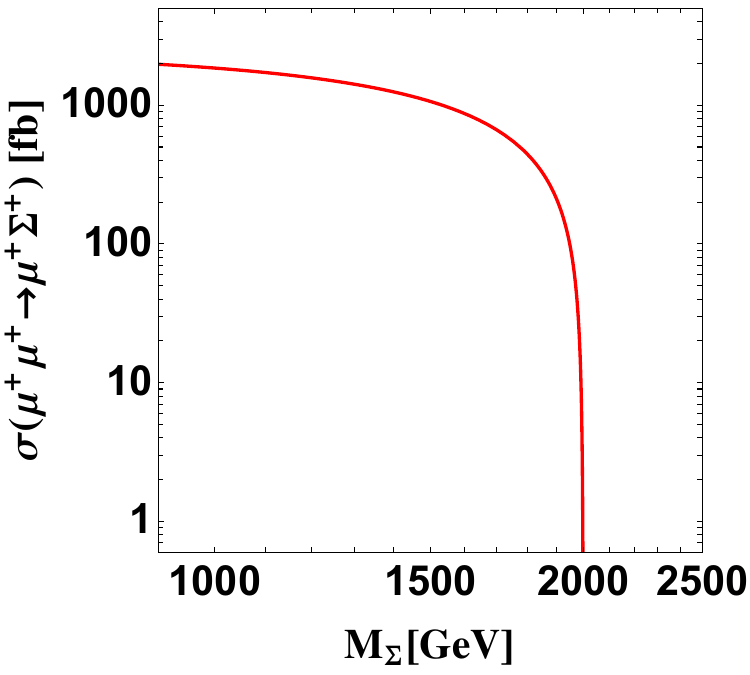}
\caption{$\Sigma^+$ production cross section as a function of triplet fermion mass in $\mu$TRISTAN experiment at $\sqrt{s}=2$ TeV center of mass energy normalized by the light-heavy mixing squared.}
\label{fig11-x}
\end{figure} 
The total production cross section as a function of the triplet mass $M_{\Sigma}$ is shown in Fig.~\ref{fig11-x}. After considering BR$(\Sigma^+ \to \mu^+ Z)=25\%$ and BR$(Z\to j j)=69.9\%$ the cross section becomes 324 fb being normalized by the square of the light-heavy mixing before cuts for $\mu^+ \mu^+ jj$ final state when $M_\Sigma=1$ TeV.
\begin{figure}[]
\centering
\includegraphics[width=0.4\textwidth,angle=0]{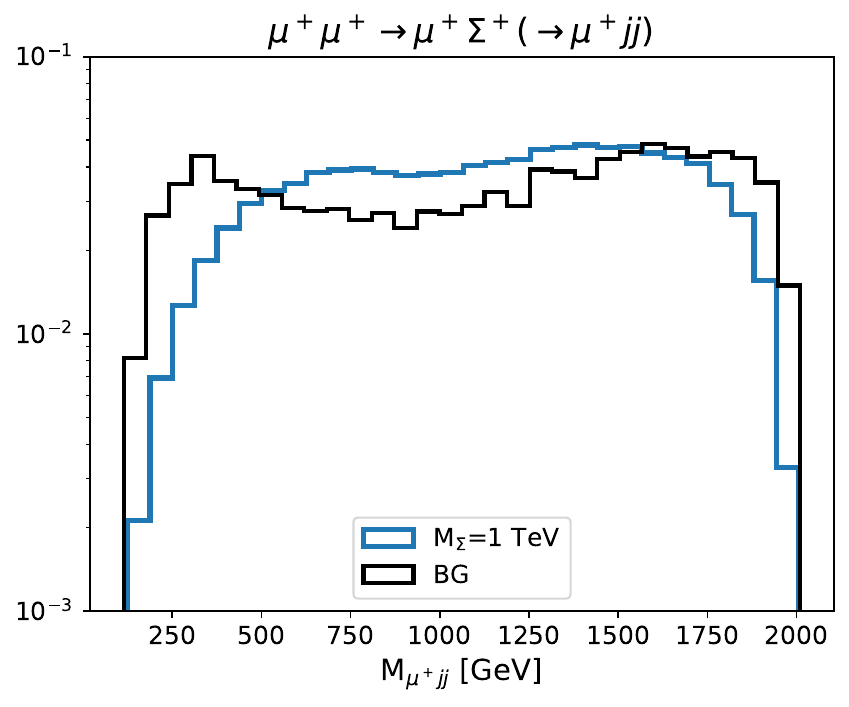}
\includegraphics[width=0.4\textwidth,angle=0]{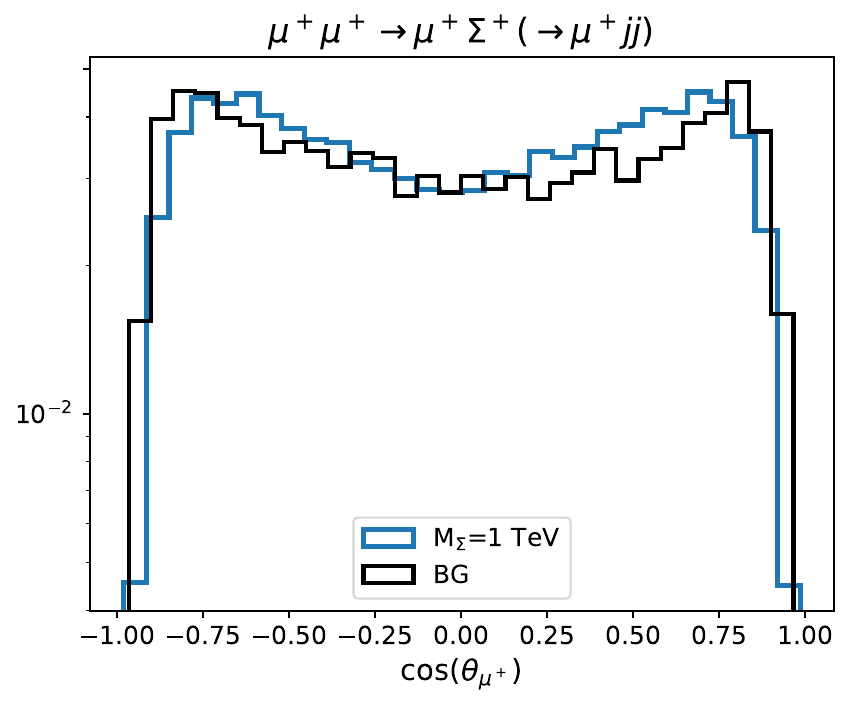} \\
\includegraphics[width=0.4\textwidth,angle=0]{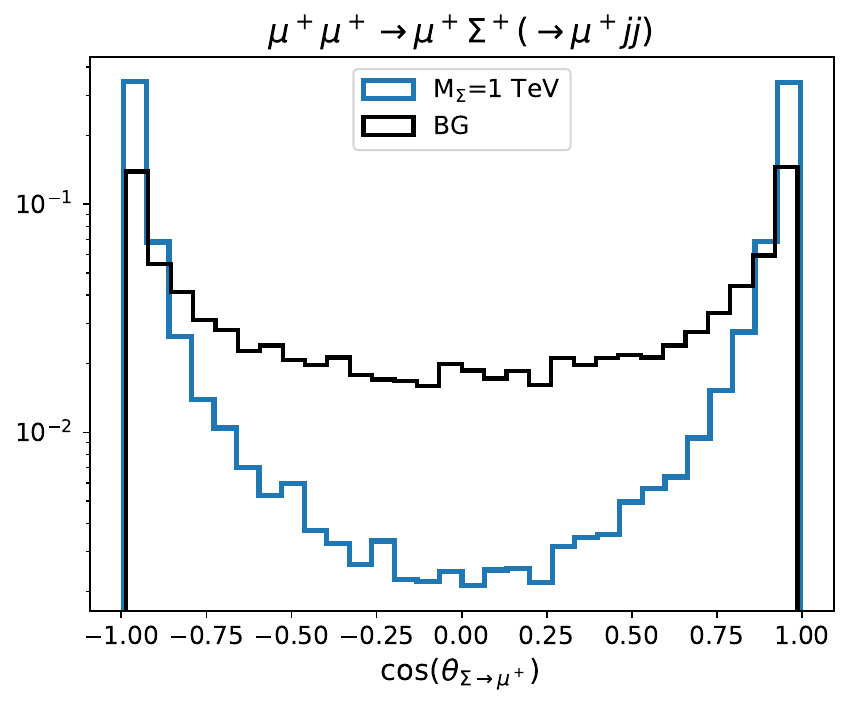}
\includegraphics[width=0.4\textwidth,angle=0]{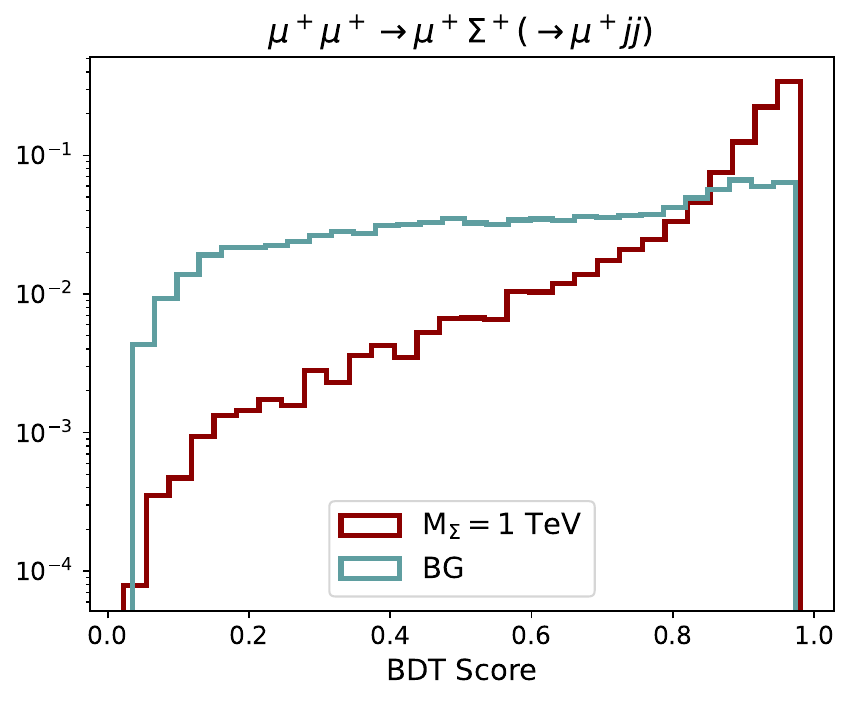}
\caption{Distributions of some kinematic variables as well as the BDT score for the signal ($\mu^+ \mu^+ \to \mu^+ \Sigma^+(\to Z \mu^+)$) and background.}
\label{typeiii_lhc}
\end{figure} 
However, as $m_\Sigma \gg m_Z +m_\mu$, the $Z$ boson is highly boosted. As a result, only the $Z$ jet can be reconstructed at the detector. The cross section of the background process $\mu^+ \mu^+ \to \mu^+ \mu^+ j j$ is 0.0341 pb. To enhance the signal-background separation, we employ a Boosted Decision Tree (BDT) algorithm trained on the following kinematic observables: 
\begin{align}
p^\mu (\mu^+), \ p^\mu(\mu_{\Sigma \to \mu}), \ p^\mu(j_Z), \ m_{\Sigma},\cos(\theta(\mu^+)), \ \cos(\theta(\mu_{\Sigma \to \mu}))~,~
\end{align}
where $p^\mu$ denotes the four-momentum and $\cos(\theta)$ specifies the polar angle (measured in the lab frame) of the corresponding particle. The reconstructed mass $m_{\Sigma}$ is derived from the vector sum of momenta of:
(i) the boosted jet $j_Z$ (identified as originating from a hadronically decaying $Z$ boson in signal processes), and
(ii) the muon $\mu_{\Sigma \to \mu}$) exhibiting minimal angular separation $\Delta R= \sqrt{(\Delta \eta)^2+(\Delta \phi)^2}$ relative to $j_Z$.
In Fig.~\ref{typeiii_lhc}, we present the distributions of some kinematic variables as well as the BDT score for our signal and background. Due to the relatively large width of the $\Sigma$ and ambiguity in combining the $\mu$ and the jet, the discrimination of signal and background is mild. Optimizing the BDT cut to maximize the signal significance, we obtain the signal 
and background (SMBG) after the cut to be 29.9 fb and 0.01633 fb respectively. Considering the light-heavy mixing squared from EWPD \cite{delAguila:2008pw,delAguila:2008cj} as $|V_{\mu \Sigma}|^2=$ 0.000289 we estimate significance $(\sigma)$ as a function of luminosity using Eq.~\ref{signi}
and show the significance as a function of luminosity in Fig.~\ref{fig12}. We find that to study $\mu^+\mu^+ jj$ final state for $\Sigma^+$ induced final state, a 2$\sigma$ significance could be obtained around 1000 fb$^{-1}$ luminosity and it could reach around 3$\sigma$ with a luminosity of 2500 fb$^{-1}$ when $M_{\Sigma}=1$ TeV. 
\begin{figure}[h]
\centering
\includegraphics[width=0.7\textwidth,angle=0]{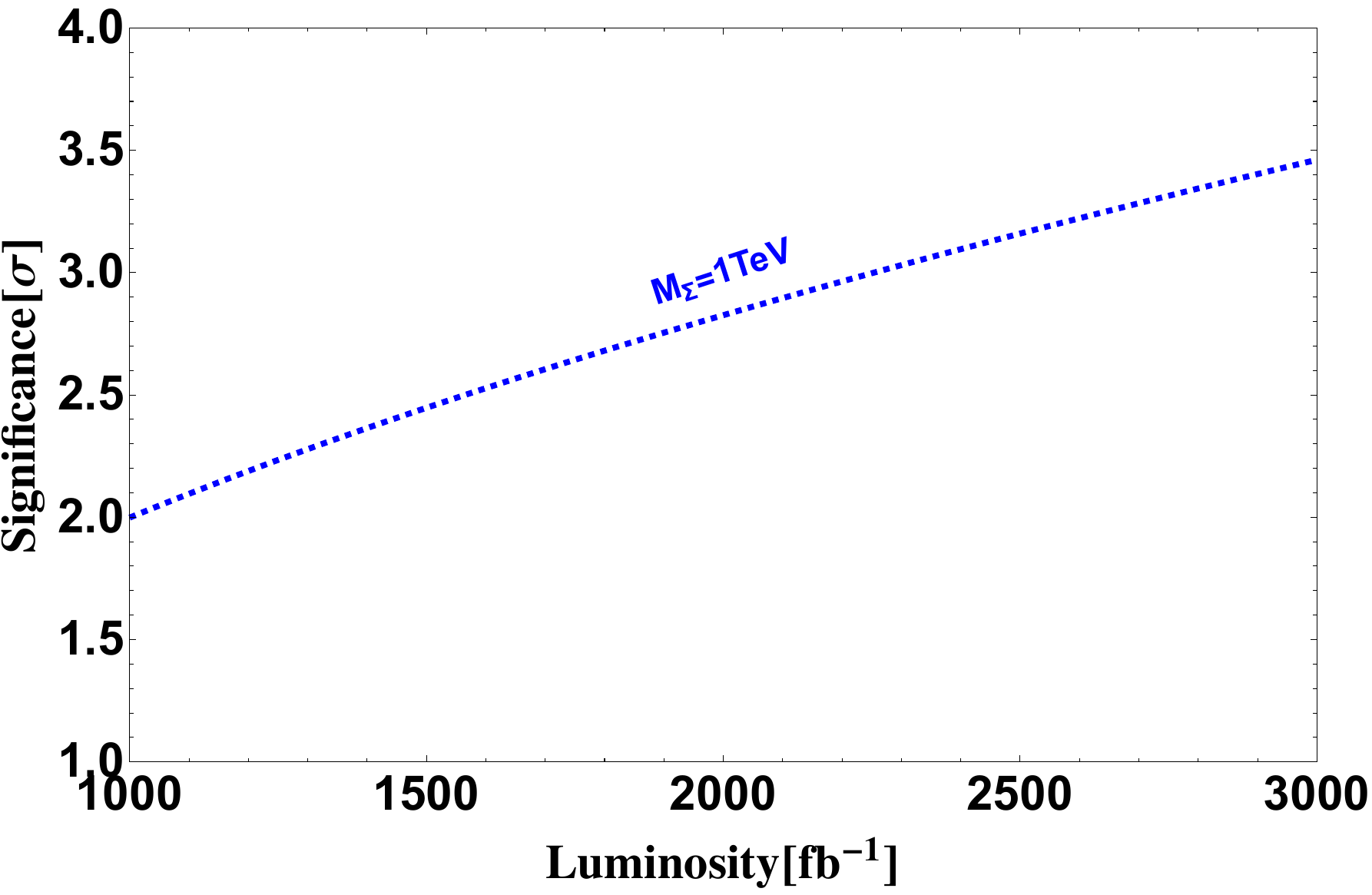}
\caption{Significance of $\Sigma^+$ production with respect to luminosity in $\mu^+ \mu^+ jj$ final state in $\mu^+ \mu^+$ collider of $\mu$TRISTAN experiment at $\sqrt{s}=2$ TeV considering $M_{\Sigma}=1$ TeV.}
\label{fig12}
\end{figure} 
\section{Conclusions}
\label{conc}
We study different tree level seesaw scenarios, namely, TeV scale type-I, type-II and type-III seesaw models at $\mu^+e^-$~($\sqrt{s}=346$ GeV) and $\mu^+\mu^+$~($\sqrt{s}=2$ TeV) collider in the proposed $\mu$TRISTAN experiment. In case of $\mu^+ e^-$ collider we study the production of heavy RHN in association with light neutrinos. These heavy neutrinos then decay through the mode of a charged leton and jets providing a single lepton, missing momentum and jets. Generating events for the signal and SM backgrounds we estimate prospective limits on the light-heavy mixing angles for 1 ab$^{-1}$ luminosity and 2$\sigma$ significance. We find that mixing $|V_{eN}|^2$ could be probed between $4.2\times 10^{-6} \leq |V_{e N}|^2 \leq 10^{-5}$ for 93 GeV $\leq M_{N} \leq 295$ GeV which is two orders of magnitude stronger than EWPD limits. Similarly the mixing $|V_{\mu N}|^2$ could be probed between $2.0 \times 10^{-6} < |V_{\mu N}|^2 \leq 10^{-5}$ for 88 GeV $\leq M_{N} \leq 304$ GeV which is also nearly two orders of magnitude stronger than the limits obtained from EWPD. We could also provide a prospective limit on $|V_{eN}^\ast V_{\mu N}|$ around $\mathcal{O}(10^{-5})$ being stronger than existing bounds form LHC for $50$ GeV $\leq M_N \leq 300$ GeV. In addition to that we have studied same sign dilepton production from the doubly charged scalar induced interactions from the type-II seesaw scenario. We found that such a scenario depends on neutrino oscillation data and hence on neutrino mass hierarchy. \ignore{Generating signal events for different final states involving photon, same sign lepton and jets we find that it could be possible to differentiate between the NO and IO.} 
More specifically, we find that two positron final state could provide nearly two orders of magnitude more events if inverted hierarchy of the neutrino mass is considered compared to the normal hierarchy scenario, whereas the dilepton final state with different flavor can not distinguish between the neutrino mass hierarchies. We also study production modes such as of $\mu^+\mu^+\to H^{++}\gamma/Z\to e^+e^+\gamma$ or $e^+e^+jj$. The cross section for this process after all kinematic cuts could make a nearly two orders of magnitude difference between normal and inverted ordering of the neutrino mass hierarchy. We find that generic SM background events could be negligibly small after the application of kinematic cuts allowing for 5$\sigma$ significance of these signals for different neutrino mass hierarchies in future. Finally we studied the left-right asymmetry from the same sign dilepton mode of mediating the doubly charged scalar boson. We find that positively charged di-muon final state could significantly deviate from the SM results and the corresponding deviations could be probed in the $\mu$TRISTAN experiment with 3$\sigma$ to 5$\sigma$ significances in future. Finally we study $\mu^+ \mu^+ jj$ final state from the production of singly charged multiplet of the triplet fermion in association with a muon for $M_{\Sigma}=1$ TeV. This final state could be observed at 2$\sigma$ significance with a luminosity of 1000 fb$^{-1}$ and the significance could reach up to a significance of 3$\sigma$ with a luminosity of 2500 fb$^{-1}$ which could be tested in $\mu$TRISTAN experiment in future.
\begin{acknowledgments}
This work was supported by the Fundamental Research Funds for the Central Universities (TN, JL), the Natural Science Foundation of Sichuan Province under grant No.~2023NSFSC1329 and the National Natural Science Foundation of China under grant No.~11905149 (JL).  The work of S.M. is supported by KIAS Individual Grants (PG086002) at Korea Institute for Advanced Study.
\end{acknowledgments}
\bibliographystyle{utphys}
\bibliography{bibliography}
\end{document}